\documentclass{article}
\usepackage{blindtext}
\usepackage{graphicx}
\usepackage{authblk}
\usepackage{subcaption}
\usepackage{xcolor}
\usepackage[fleqn]{amsmath}
\usepackage[font=small,skip=3pt]{caption}
\usepackage{makecell}
\usepackage{hyperref} 
\usepackage{cleveref}
\usepackage{url}
\usepackage{amssymb}
\usepackage[style=phys,
    backend=bibtex, 
    natbib=false,
    style=phys,
    biblabel=brackets,
    giveninits=false,
    abbreviate=false,
    doi=false,
    url=false,
    isbn=false,
    giveninits=true,
    block=space,
    minbibnames=3, maxbibnames=3,
    backref=false, backrefstyle=two,]{biblatex}
\DeclareFieldFormat{apacase}{#1}
\AtEveryBibitem{\clearfield{pages}} 
\AtEveryBibitem{\clearfield{doi}} 
\author[1,*]{\normalsize J. F. Parisi}
\author[2]{A. O. Nelson} 
\author[3,1]{W. Guttenfelder}
\author[4]{R. Gaur}
\author[1]{J. W. Berkery}
\author[1]{S. M. Kaye}
\author[5]{K. Barada}
\author[6]{C. Clauser}
\author[1]{A. Diallo}
\author[7,8]{D. R. Hatch}
\author[1]{A. Kleiner} 
\author[1]{M. Lampert}
\author[5]{T. Macwan}
\author[1]{J. E. Menard}
\affil[1]{\small Princeton Plasma Physics Laboratory, Princeton University, Princeton, NJ, USA}
\affil[2]{Department of Applied Physics and Applied Mathematics, Columbia University, New York, NY, USA}
\affil[3]{Type One Energy, 8383 Greenway Boulevard, Middleton, WI, USA}
\affil[4]{Department of Mechanical and Aerospace Engineering, Princeton University, Princeton, NJ, USA}
\affil[5]{University of California, Los Angeles, Los Angeles, CA, USA}
\affil[6]{Plasma Science and Fusion Center, Massachusetts Institute of Technology, Cambridge, MA, USA}
\affil[7]{Institute for Fusion Studies, University of Texas at Austin, Austin, Texas, USA}
\affil[8]{ExoFusion, Austin, Texas, USA}

\title{\LARGE Stability and Transport of Gyrokinetic Critical Pedestals}
\date{\vspace{-5ex}}

\begin{document}

\maketitle

\href{mailto: jparisi@pppl.gov}{*Email: jparisi@pppl.gov}

\vspace{10pt}

\begin{abstract}
A gyrokinetic threshold model for pedestal width-height scaling prediction is applied to multiple devices and to a shaping and aspect-ratio scan giving $\Delta_{\mathrm{ped}} = 0.92 A^{1.04} \kappa^{-1.24} 0.38^{\delta} \beta_{\theta,\mathrm{ped}}^{1.05}$ for pedestal width $\Delta_{\mathrm{ped}}$, aspect-ratio $A$, elongation $\kappa$, triangularity $\delta$, and normalized pedestal height $\beta_{\theta,\mathrm{ped}}$. We also find a width-transport scaling $\Delta_{\mathrm{ped} } = 0.028 \left(q_e/\Gamma_e - 1.7 \right)^{1.5} \sim \eta_e ^{1.5}$ where $q_e$ and $\Gamma_e$ are turbulent electron heat and particle fluxes and $\eta_e = \nabla \ln T_e / \nabla \ln n_e$ for electron temperature $T_e$ and density $n_e$. Pedestals close to those limited by kinetic-ballooning-modes (KBMs) have modified turbulent transport properties compared to strongly driven KBMs. The role of flow shear is studied as a width-height scaling constraint and pedestal saturation mechanism for a standard and wide pedestal discharge.
\end{abstract}
\section{Introduction} \label{sec:1}

The H-mode pedestal is an edge transport barrier that forms when a strongly heated tokamak plasma transitions into a high confinement regime \cite{Wagner1982,Kaye1984}. Due to a significant improvement in confinement, H-mode is a leading candidate for burning plasma scenarios \cite{Shimada2007,Kessel2009,Kinsey2011,Garofalo2015,Snyder2019,Creely2020}. The pedestal pressure profile has a characteristic width and height \cite{Mahdavi2003}, whose values can be found with the EPED stability threshold model \cite{Snyder2009}, often with remarkable success \cite{Snyder2011, Walk2012, Snyder2015, Hughes2018, Snyder2019} and some exceptions \cite{Diallo2013,Saarelma2017}. EPED combines a local gradient constraint controlled by infinite-$n$ ideal-ballooning-mode (IBM) stability \cite{Connor1979} and a macroscopic stability constraint controlled by peeling-ballooning-mode (PBM) stability \cite{Snyder2002}; the intersection of IBM and PBM constraints gives a pedestal width-height prediction.

However, extra information from pedestal models is required in order to understand current experiments and design future devices \cite{Hughes2020,Guttenfelder2021}. For example, while IBM and PBM stability might suffice to predict the pedestal pressure's width-height trajectory, the separate evolution of density and temperature pedestal profiles is needed to determine plasma heating and fueling sources that are consistent with such profiles. Given that gyrokinetic instabilities are often sensitive to density and/or temperature gradients and their relative sizes \cite{Cowley1991,Jenko2001}, non-IBM transport mechanisms are expected to play a significant role in pedestal evolution \cite{Wang2012,Hatch2016,Hatch2017,Churchill2017,Chapman2022,Field2023}. Equilibrium flow shear, part of which is generated by the temperature and density profiles themselves, is also important in pedestal formation \cite{Groebner1990} and the inter-ELM cycle \cite{Schirmer2006}, and is hypothesized as necessary at the pedestal top to allow the pedestal to widen \cite{Snyder2009} by stabilizing ion-temperature-gradient and trapped-electron-mode instabilities. Given the importance of flow shear in the pedestal -- and also its pessimistic scaling with $\rho_i / a$ to future devices \cite{Kotschenreuther2017}, where $\rho_i$ is the ion gyroradius and $a$ the minor radius -- generating sufficient flow shear in future reactors (or developing alternative methods \cite{Kotschenreuther2023}) is an important task.
 
The pedestal is also highly coupled to the core and scrape-off layer (SOL) \cite{Meyer2017,Rodriguez2022b,Saarelma2023}, meaning that the pedestal's evolution cannot be considered in isolation. Solving this coupling in sufficient detail remains one of the biggest challenges in fusion research \cite{community2020} due in part to the high dimensionality of the problem, the complex interactions between phenomena, and the uncertainties in models and measurements. For example, predicting the pedestal density profile is a very hard problem. The density pedestal in current experiments is sensitive to neutral particles ionizing in the pedestal and to transport processes \cite{Groebner2002,Kotschenreuther2019,Mordijck2020,Saarelma2023,Kit2023,Horvath2023}. Changes in divertor and SOL physics and the neutral ionization source in the pedestal will change the pedestal pressure and current profile, in turn strongly affecting core physics, which in turns affects the pedestal, and so on. The strong coupling of the pedestal to other regions motivates high-fidelity models to prevent large error propagation.

In this work, we focus on three challenges mentioned above: (1) accuracy of the pedestal width-height prediction, (2) pedestal heat and particle transport constraints, and (3) flow shear at the pedestal top. 

For (1), we introduce a linear gyrokinetic threshold model that builds on the EPED Ballooning Critical Pedestal model \cite{Snyder2009} to predict the width-height pedestal scaling across aspect-ratio, shaping, and devices. We call these width-height scaling expressions Gyrokinetic Critical Pedestal constraints. We perform a study of how shaping and aspect-ratio enter the gyrokinetic pedestal width-height scaling, which may explain the differences in width-height experimental measurements between most devices and the National Spherical Torus Experiment (NSTX) \cite{Snyder2009, Diallo2013, Parisi2023_ARXIV}, with NSTX typically featuring much wider pedestals. Varying the pedestal shaping and aspect-ratio may also provide new opportunities for controlling pedestal width and height in experiments. For (2), we study heat and particle transport around the gyrokinetic width-height scalings and find the dependence of pedestal width on the electron heat to particle transport ratio. For (3), we study flow shear at the pedestal top in a regular NSTX ELMy H-mode and a NSTX ultra-wide enhanced pedestal (EP) H-mode \cite{Maingi2010}.

The paper layout is as follows: in \Cref{sec:workflow}, we describe the process for varying the pedestal width and height with self-consistent equilibrium reconstruction. We introduce the gyrokinetic and ideal-ballooning framework used for studying these equilibria. A linear threshold model for gyrokinetic and ideal-ballooning stability is then applied to several devices in \Cref{sec:3} to find width-height scaling expressions. The shaping and aspect-ratio dependence of the pedestal width-height-shaping scaling is explored in \Cref{sec:4}. In \Cref{sec:transport}, we study gyrokinetic turbulent transport properties close to the Gyrokinetic Critical Pedestal boundaries and find scaling expressions that relate the pedestal width and turbulent transport ratios. In \Cref{sec:flowshear} we explore the role of flow shear as an additional width-height constraint. In \Cref{sec:combined_constraints}, we combine the stability, flow shear, and transport constraints to determine regions of pedestal width-height space that are accessible.

\section{Width-Height Scaling Workflow} \label{sec:workflow}

\begin{figure}[tb]
 \centering
    \includegraphics[width=0.7\textwidth]{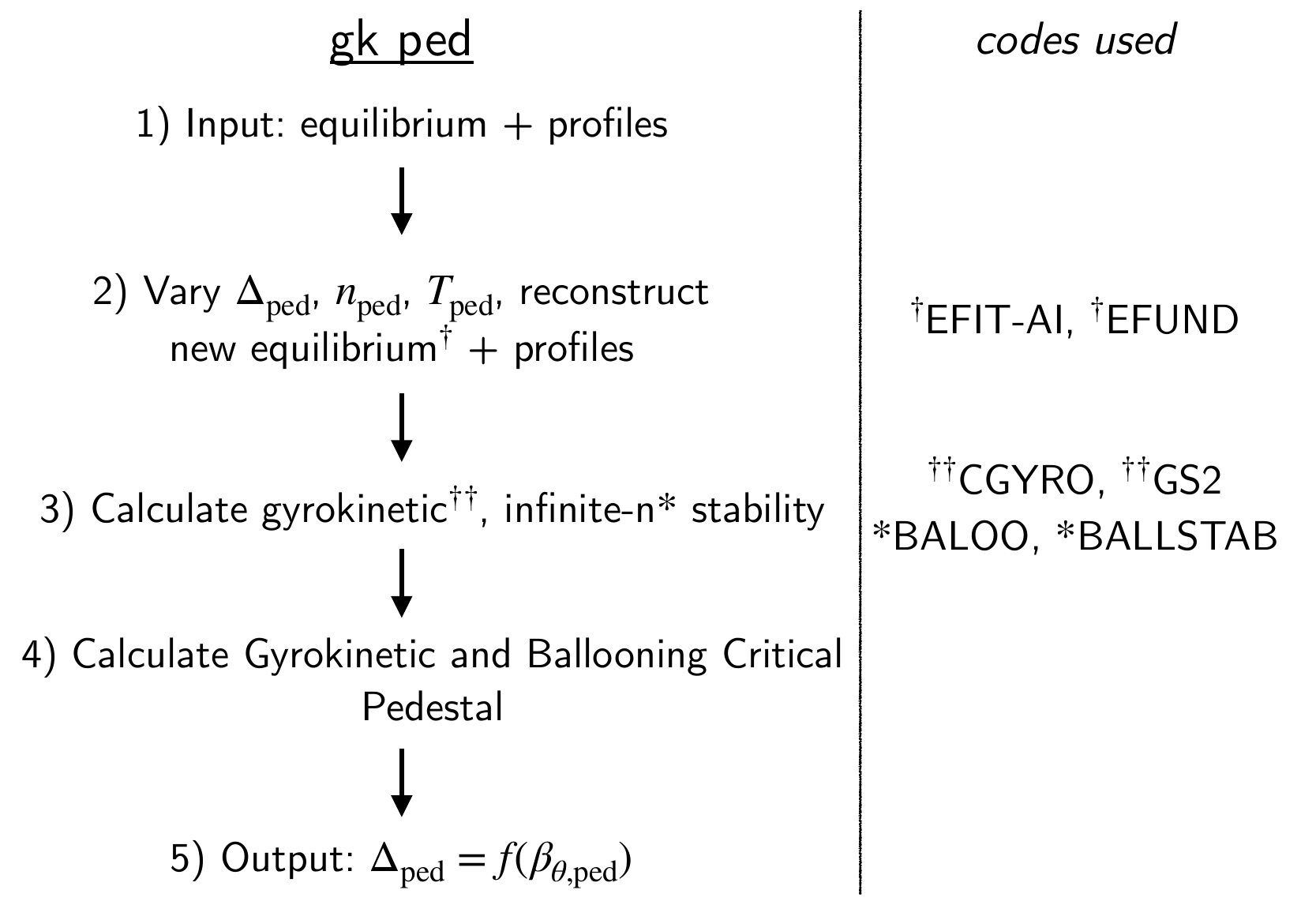}
        \caption{gk\_ped workflow.}
    \label{fig:gkpedworkflow}
\end{figure}

In this section, we describe the workflow to compute the width-height scaling using the new framework gk\_ped, outlined in \Cref{fig:gkpedworkflow}. Currently, gk\_ped is implemented as a module in the integrated modeling and data analysis software OMFIT \cite{OMFIT2015}. In addition to the OMFIT tools, we use EFIT-AI \cite{Lao2022} combined with EFUND \cite{Appel2006} for equilibrium reconstruction, GS2 \cite{Dorland2000,Barnes2021} and CGYRO \cite{Candy2016} for gyrokinetic simulations, and BALOO \cite{Miller1997} and ball\_stab \cite{Barnes2021,Gaur2023a} for infinite-n ballooning simulations.

\subsection{Equilibrium Variation and Reconstruction} \label{sec:2}

The first step in gk\_ped is to input an equilibrium and profiles. The electron temperature is parameterised as 
\begin{equation}
\begin{aligned}
T_e(\psi) = & T_{e,\mathrm{c}} \mathrm{H} \left[  \psi_{\mathrm{ped,T_e}} -  \psi \right] (1-\psi^{\alpha_{T_1}})^{\alpha_{T_2}} \\ 
& + T_{e0} \left( \tanh(2) - \tanh \left(\frac{\psi-\psi_{\mathrm{mid,T_e}}}{ \Delta_{T_e}/2} \right) \right) + T_{\mathrm{e,sep}},
\label{eq:2}
\end{aligned}
\end{equation}
with the same functional form for electron density $n_e$ \cite{Mahdavi2003,Snyder2009}. Here, H is a step function, $\psi$ is the poloidal flux normalized to 0 at the magnetic axis and 1 at the last-closed-flux-surface, $T_{e,\mathrm{c}}$, $T_{e0}$, $\alpha_{T_{1,2}}$ are constants, $\Delta_{T_e}$ and $\Delta_{n_e}$ are the electron temperature and density pedestal widths. The pedestal heights $T_{e,\mathrm{ped}}$ and $n_{e,\mathrm{ped}}$ are $T_{e,\mathrm{ped}} = T_e(\psi_{\mathrm{ped,T_e}}), \;\; n_{e,\mathrm{ped}} = n (\psi_{\mathrm{ped,n_e}})$ where $T_{e,\mathrm{sep}}$ and $n_{e,\mathrm{sep}}$ are evaluated at $\psi =1$, and $\psi_{\mathrm{ped,T_e}} = \psi_{\mathrm{mid,T_e}} - \Delta_{T_e} / 2$,  $\psi_{\mathrm{ped,n_e}} = \psi_{\mathrm{mid,n_e}} - \Delta_{n_e} / 2$, For the ions, the density profiles are given by quasineutrality and the temperature profiles by enforcing constant $T_i / T_e$ across the radial profile. The current density $J(\psi)$ is the sum of bootstrap $J_{\mathrm{bs}}$ and non-bootstrap $J_{\mathrm{non-bs} }$ contributions,
\begin{equation}
J = k J_{\mathrm{bs}} + J_{\mathrm{non-bs} }.
\end{equation}
We vary the pedestal width and height with constant plasma current $I_p$, constant $\beta_N = \beta_T a B_{T0} / I_p$, and self-consistent bootstrap current. Here, $\beta_T = 8\pi \langle p \rangle_V / B_{T0}^2 $, where $\langle p \rangle_V$ is the pressure averaged over the plasma volume and $B_{T0}$ is the magnetic field strength at the magnetic axis. In order to keep $I_p$ constant due to changing $J_{\mathrm{bs}}$ resulting from different profiles, $J_{\mathrm{non-bs} }$ (obtained from the original input equilibrium), is multiplied by a constant $k$. The bootstrap current can be calculated from several formulae in the literature \cite{Sauter1999,Redl2021} that are implemented in OMFIT, but is recommended that it be the same formula that generated the original input equilibrium. The quantity $\beta_N$ is kept constant by an iterative procedure that allows $\beta_N$ to vary from the input equilibrium by at most $1\%$.

Once the user is satisfied with the profile parameterization, the pedestal width and height are varied. The height can be varied in two ways: at constant $T_{e,\mathrm{ped}}$ with varying $n_{e,\mathrm{ped}}$, or at constant $n_{e,\mathrm{ped}}$ with varying  $T_{e,\mathrm{ped}}$. As a default option, $n_{e,\mathrm{ped} }/ n_{e,\mathrm{sep} }$ is held constant when varying $n_{e,\mathrm{ped}}$ and $T_{e,\mathrm{sep}}$ is held constant when varying $T_{e,\mathrm{ped}}$. The total pedestal width is 
\begin{equation}
\Delta_{\mathrm{ped}} = (\Delta_{n_e} + \Delta_{T_e})/2,
\end{equation}
the pedestal top location is $ \psi_{\mathrm{ped}} = \psi_{\mathrm{mid}} - \Delta_{\mathrm{ped}} / 2$ where $\psi_{\mathrm{mid}} =  (\psi_{\mathrm{mid,n_e}}+ \psi_{\mathrm{mid,T_e}})/2$, and the normalized pedestal height is
\begin{equation}
\beta_{\theta, \mathrm{ped}} = 8 \pi p_{\mathrm{ped}} /\overline{B}_{\mathrm{pol}}^2,
\end{equation}
where $p_{\mathrm{ped}} = 2 p_e (\psi = \psi_{\mathrm{ped}} )$ and $\overline{B}_{\mathrm{pol}} =4\pi I_p /  l c $ with flux surface circumference $l$ and speed of light $c$ \cite{Snyder2009,Smith2022,Parisi2023_ARXIV}.

\subsection{Gyrokinetic and Ideal Analysis} \label{sec:2b}

In this section, we describe briefly the gyrokinetic and ideal simulations performed in the pedestal, and the gyrokinetic mode identification scheme.

We use GS2 \cite{Barnes2021} and CGYRO \cite{Candy2016} to solve the electromagnetic gyrokinetic equation \cite{Catto1978,Frieman1982,Parra2008,Abel2013} for distribution function $g_s = \delta f_s + Z_s e  F_{0s} (\phi - \chi) /T_s $,
\begin{equation}
    \partial g_s / \partial t + \mathbf{\dot{R}_s} \cdot \nabla \left[ g_s + (Z_s e \chi_s / T_s) F_{0s} \right] + v_{\chi,s} \cdot \nabla F_{0s} - C_s = 0, 
    \label{eq:gke}
\end{equation}
and Maxwell's equations. Here, $\delta f_s$ is the total turbulent distribution function, $F_{0s}$ is a Maxwellian, $\mathbf{\dot{R}}$ is the guiding-center particle drift, $\phi$ and $\chi_s$ are the electrostatic and gyroaveraged gyrokinetic potentials \cite{Abel2013}, $v_{\chi,s} = c \mathbf{B} \times \nabla \chi_s/B^2$, $\mathbf{B}$ is the magnetic field, and $C_s$ is a collision operator. We also use BALOO \cite{Miller1997} and ball\_stab \cite{Barnes2021,Gaur2023a} to solve the infinite-n ballooning equation for the ballooning eigenfunction $Y$ and frequency $\omega$ \cite{Connor1979}, $\omega^2 \lambda Y = - d/d \theta [ g \; d Y/d \theta ] + u Y $ where $g$, $u$, and $\lambda$ are geometric coefficients \cite{Gaur2023}. We typically solve both ideal and gyrokinetic equations in the pedestal width-height model.

Linear gyrokinetic simulations are performed for a range of binormal wavenumbers $k_y$, typically at scales where kinetic-ballooning-modes (KBMs) \cite{Tang1980,Aleynikova2018} are expected to be most virulent, i.e. $k_y \rho_i \approx 0.05 - 0.2$, and in a radial domain across the pedestal that is evenly spaced in $\psi$. Here, $\rho_i$ is the ion gyroradius. Once simulations are converged, a mode finder routine adopting a fingerprints-like \cite{Kotschenreuther2019} approach is used to identify the mode type. To aid mode identification, each gyrokinetic simulation is launched with a second simulation with a slightly increased $\beta$ value (where $d \beta / d \psi$ is unchanged) --- this is particularly helpful for distinguishing between KBMs and trapped-electron-modes (TEMs) \cite{Ernst2004, Hatch2015}, since both KBMs and TEMs often have similar transport characteristics \cite{Kotschenreuther2019,Clauser2022}. In \Cref{tab:tab1}, we describe the criteria for each gyrokinetic mode. In addition to KBMs and TEMs, we consider micro-tearing modes (MTMs) \cite{Drake1980,Connor1990,Hatch2021,Hardman2022}, electron-temperature-gradient (ETG) modes \cite{Dorland2000, Jenko2001, Parisi2020, Adkins2022}, tearing ETG (TETG) modes \cite{Zocco2015}, and ion-temperature-gradient (ITG) modes \cite{Rudakov1961,Hammett1990,Cowley1991,Nunami2011}. In \Cref{tab:tab1}, $\chi_s$ and $D_s$ are the turbulent heat and particle diffusivities (later defined in \Cref{eq:q_and_gamma}), $\mathcal{P}(A_{\parallel }) = 1- |\int A_{\parallel} d\theta| / \int | A_{\parallel}| d\theta$ \cite{Hatch2012} is the mode parity in the fluctuating field $A_{\parallel}$, $\gamma$ is the linear growth rate, and $\omega_R$ is the real frequency.

\begin{table}
\caption{\label{tab:example} Mode finder criteria in gk\_ped. A `$-$' indicates that the quantity is not considered in the mode criterion.} %
  \begin{tabular}{|| c | ccccccc }
    \hline
    Mode & $\chi_i / \chi_e$ & $D_e / \chi_e$ & $D_i / \chi_i$ & $D_e / (\chi_e + \chi_i)$ & $\mathcal{P}(A _{\parallel })$ & $\partial \gamma / \partial \beta$ & $\omega_R$ \\
    \hline
    KBM & $\sim1$ & $\sim1$ & $\sim1$ & $\gtrsim 1/3$ & $ 1$ & $>0$ & $-$ \\ \hline
    TEM & $\sim 1$ & $\sim1$  &  $-$ & $\lesssim 1/3$ & $ 1$ & $ <0$ & $-$ \\ \hline
    MTM & $\ll 1$ & $\ll 1$  & $-$ & $-$ & $<1$ & $>0$ & $\simeq \omega_{*e}$ \\ \hline
    ETG & $\ll 1$ & $\ll 1$ & $-$ & $-$ & $ 1$ & $-$ & $-$ \\ \hline
    TETG & $\ll 1$ & $\ll 1$ & $-$ & $-$ & $< 1$ & $-$ & $-$ \\ \hline
    ITG & $\gg 1$ & $-$  & $\ll 1$ & $-$ & $1$ & $ <0$ & $-$ \\ \hline
  \end{tabular}
\label{tab:tab1}
\end{table}

Because of the wide range of devices we study and the often unusual nature of gyrokinetic instabilities in the pedestal, many of the mode criteria quantities in \Cref{tab:tab1} are not hard cutoffs, but rather, approximate values that the mode should satisfy -- the more criteria that a given instability satisfies, the more confidence in a given mode type identification. In our experience, distinguishing between KBM, TEM, and ITG was often challenging and required using $D_e / (\chi_e + \chi_i)$ and $\partial \gamma / \partial \beta$. With the exception of the MTM, we ignore the mode's real frequency in automated mode identification. For all gyrokinetic and ideal simulations in this paper, we only study the radial wavenumber $k_x = 0$ because we focus on KBM stability. While $k_x = 0$ is typically the most unstable mode for KBM \cite{Kennedy2023p}, for other instabilities this may not be the case \cite{Told2008,Angioni2017,Hatch2019,Parisi2020,Hardman2023,Leppin2023}.

We also perform ideal infinite-n ballooning simulations at the same radial locations as the gyrokinetic simulations. At each flux-surface, there are three stability states for the ideal mode: first-stable, unstable, and second-stable.

\subsection{The Ballooning and Gyrokinetic Critical Pedestal} \label{sec:2c}

\begin{figure*}[tb]
    \centering
    \begin{subfigure}[t]{0.4\textwidth}
        \centering
        \includegraphics[width=\textwidth]{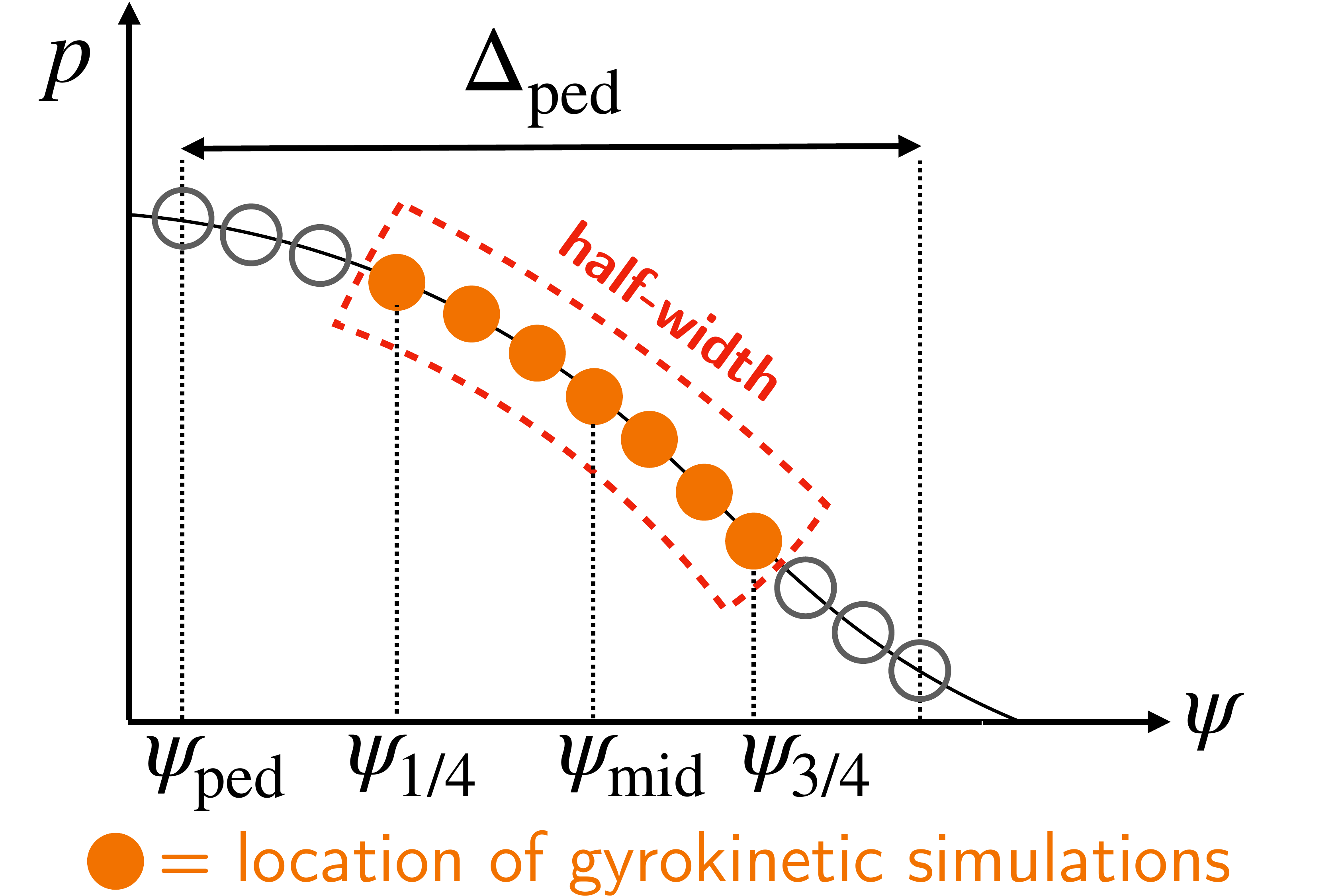}
        \caption{Locations of gyrokinetic simulations in pedestal half-width}
    \end{subfigure}
    ~
    \begin{subfigure}[t]{0.55\textwidth}
        \centering
        \includegraphics[width=\textwidth]{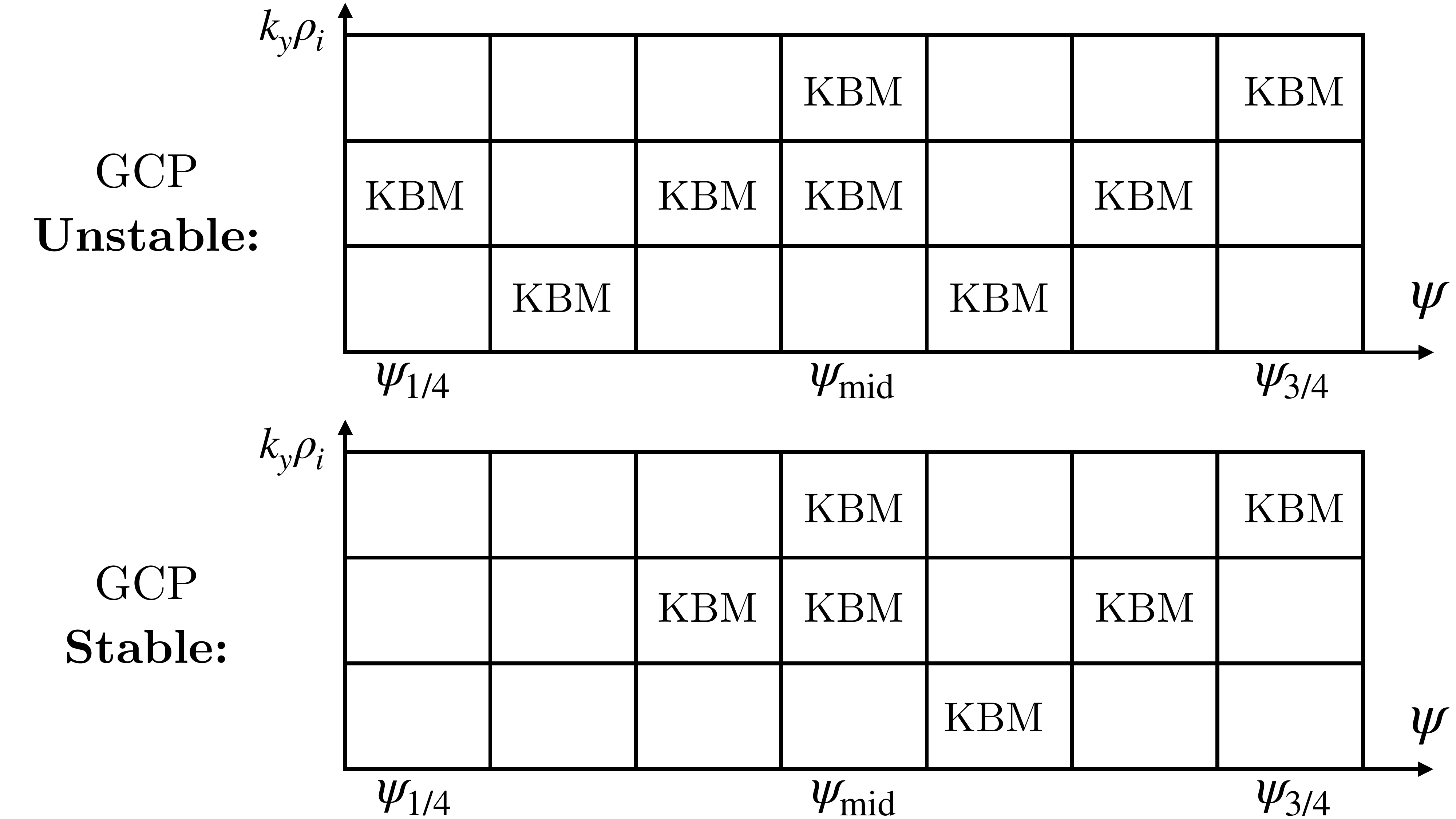}
        \caption{GCP calculation with dominant gyrokinetic instability versus binormal wavenumber $k_y \rho_i$ and radial location $\psi$.}
    \end{subfigure}
    \caption{Schematics of (a) gyrokinetic simulation radial locations and (b) Gyrokinetic Critical Pedestal (GCP) calculation with dominant gyrokinetic instability versus binormal wavenumber $k_y \rho_i$ and radial location $\psi$. Top: GCP unstable example since KBM unstable for at least a single $k_y \rho_i$ value at each radial location. Bottom: GCP stable location. Only KBM stability status is shown.}
    \label{fig:radial_half_width}
\end{figure*}

In this section, we describe the calculation of the Gyrokinetic Critical Pedestal (GCP) and Ballooning Critical Pedestal (BCP) \cite{Snyder2009} using information from gyrokinetic and ideal infinite-n simulations.

Extensive experimental and theoretical work has shown that the pedestal pressure gradient is often limited by ballooning modes \cite{Snyder2009,Groebner2010,Wan2012,Hughes2013,Merle2017}. In the EPED model, the ideal infinite-$n$ mode is hypothesized to set the steepest pressure gradient a pedestal achieves across the pedestal half-width. If every radial location within the half-width is ideal-ballooning unstable, EPED determines that the pedestal profile is no longer physically accessible, a constraint known as the Ballooning Critical Pedestal (BCP) \cite{Snyder2009}. The EPED model has been applied successfully to multiple experiments \cite{Snyder2011, Walk2012, Snyder2015, Hughes2018, Snyder2019}. However, recent work \cite{Parisi2023_ARXIV} showed that for ELMy NSTX pedestals, kinetic-ballooning -- rather than ideal-ballooning -- stability is needed to match width-height scalings with experiment \cite{Diallo2013}. Such a constraint using the KBM stability threshold is called the Gyrokinetic Critical Pedestal (GCP) \cite{Parisi2023_ARXIV,Parisi_arxiv2023_2,Berkery2024}. The pedestal half-width region used for the BCP and GCP is shown schematically in \Cref{fig:radial_half_width}(a). For the GCP calculation, if KBM is unstable at any $k_y \rho_i$ wavenumber for a given radius, that radius counts as `unstable' toward the GCP calculation. If all radii within the pedestal half-width are `unstable,' the pedestal is GCP unstable. A GCP unstable and stable pedestal are shown schematically in \Cref{fig:radial_half_width}(b). 

Practically, to find the BCP and GCP we start from an input equilibrium, typically calculated from an experiment, and construct a set of equilibria with varied pedestal width and height as outlined in \Cref{sec:2}. We then evaluate ideal-ballooning and gyrokinetic stability across the pedestal half-width on all of these equilibria, and find the boundary in $\Delta_{\mathrm{ped}}$, $\beta_{\theta,\mathrm{ped}}$ coordinates between equilibria that are accessible and inaccessible according to the BCP and GCP. A width-height scaling is found by fitting this boundary to a function with constants $C$ and $G$
\begin{equation}
    \Delta_{\mathrm{ped}} = C \left(\beta_{\theta,\mathrm{ped}}\right)^{G}.
    \label{eq:delta_fit}
\end{equation}
In this paper we describe the two cases of varying $\beta_{\theta,\mathrm{ped}}$ at fixed $n_{e,\mathrm{ped}}$ and fixed $T_{e,\mathrm{ped}}$, which previous work \cite{Parisi2023_ARXIV} has shown gave significant differences in the width-height scaling.

\subsection{Pedestal Bifurcation}

 \begin{figure}[tb]
    \centering
    \includegraphics[width=0.42\textwidth]{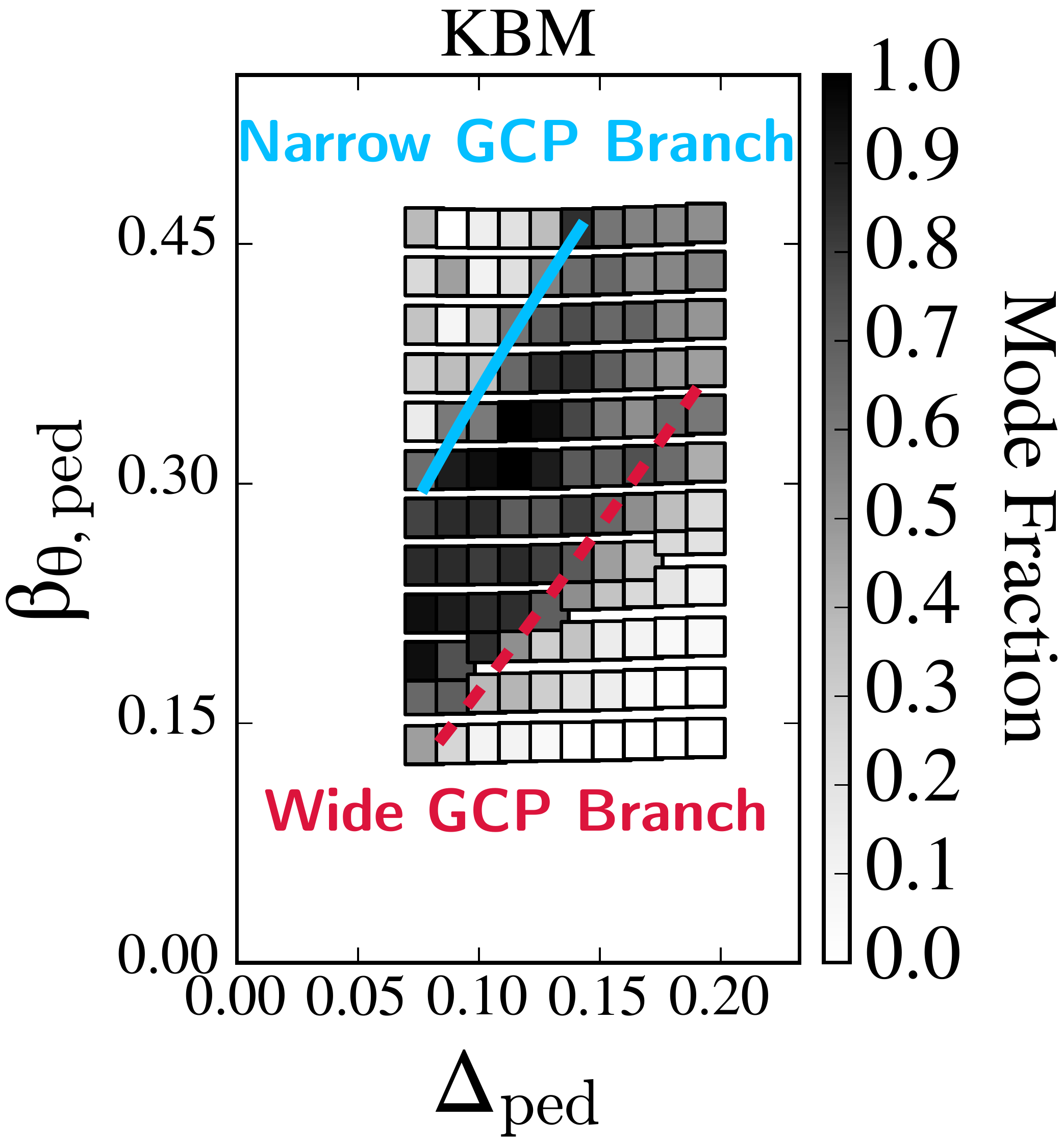}
        \caption{Fraction of KBM dominance across the pedestal half-width for fixed $n_{e,\mathrm{ped}}$ in NSTX 139047, with wide and narrow GCP branches. The mode fraction is the fraction of all linear gyrokinetic simulations across the pedestal half-width and binormal wavenumbers (here $k_y \rho_i \in [0.06, 0.12, 0.18]$) where the fastest growing mode is classified as KBM, MTM, etc.}
    \label{fig:ped_bif_outline}
\end{figure}

Recently, it was shown that a bifurcation in KBM stability caused by aspect-ratio and shaping might be responsible for the variation in width-height scaling across devices \cite{Parisi_arxiv2023_2}. If both first and second KBM stability can be accessed robustly across the pedestal half-width at different widths and heights, this led to two solutions for $\Delta_{\mathrm{ped}}$: a wide and narrow GCP branch. In \Cref{fig:ped_bif_outline}, we show an example of an equilibrium where both the wide and narrow GCP branches exist. The mode fraction in \Cref{fig:ped_bif_outline} corresponds to the fraction of all modes across the pedestal half-width and simulated $k_y \rho_i$ wavenumbers that are KBMs. In \Cref{fig:ped_bif_outline}, three binormal wavenumbers are used $k_y \rho_i \in [0.06, 0.12, 0.18]$, so the minimum KBM mode fraction required to trigger the wide or narrow GCP is 1/3. Throughout this paper, we will refer frequently to the wide and narrow GCP branches.

It has also been demonstrated that there is a bifurcation in the macroscopic constraint for pedestal prediction, peeling-ballooning-mode (PBM) stability. First and second PBM stability was achieved in DIII-D by higher fueling and strong positive triangularity \cite{Snyder2015,Snyder2019} and in TCV by varying the triangularity from negative to positive \cite{Merle2017}. 

\section{Device Scan} \label{sec:3}

\begin{figure}[tb]
 \centering
    \includegraphics[width=0.6\textwidth]{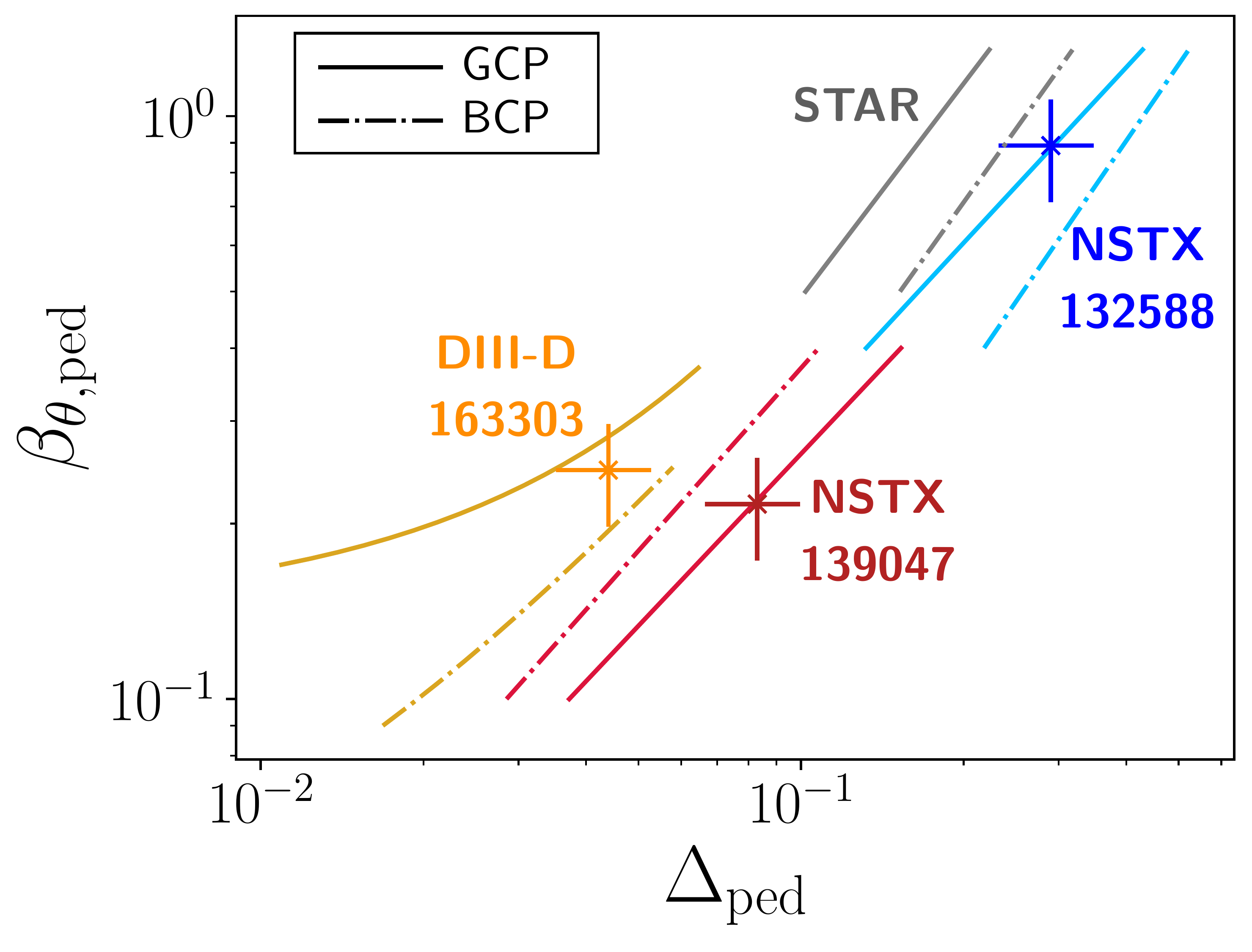}
        \caption{Gyrokinetic Critical Pedestal (GCP) and Ballooning Critical Pedestal (BCP) scaling expressions with experimental points indicated by markers and uncertainty bars of 20 \%. The design point for STAR has not yet been finalized.}
    \label{fig:phasebcpgcp}
\end{figure}

In this section, we give examples of the GCP and BCP calculations for NSTX, DIII-D, and STAR devices. The GCP for more devices is given in \cite{Parisi_arxiv2023_2}.

\subsection{NSTX}

We study two NSTX discharges: NSTX 139047 is an ELMy NSTX H-mode \cite{Diallo2013}, and NSTX 132588 \cite{Maingi2017} is a ultra-wide-pedestal lithiated enhanced-pedestal (EP) H-mode. In \Cref{fig:phasebcpgcp} we plot the GCP with solid lines and the BCP with dash-dotted lines. For both NSTX 139047 and NSTX 132588, the width-height scaling expression is in excellent agreement with the experimental point, strong evidence that KBM is limiting the pedestal width and height. Notably, for NSTX 139047, the GCP gives a less steep pedestal than the BCP, indicating that the pedestal is limited by KBM first-stability, and therefore is the wide GCP. In contrast, for NSTX 132588, the GCP gives a steeper pedestal than the BCP, indicating that the pedestal is in KBM second-stability and is therefore the narrow GCP. %

\subsection{DIII-D}

We now find the BCP and GCP for DIII-D 163303 \cite{Grierson2018}, a previously published ELMy H-mode discharge that is used to study wall conditions, the L-H transition power threshold, and outgassing. Shown in \Cref{fig:phasebcpgcp}, the experimental point for this equilibrium is at a $\beta_{\theta,\mathrm{ped}}$ value slightly above the ballooning critical pedestal but slightly below the gyrokinetic critical pedestal. However, within 20\% uncertainty, this equilibrium is consistent with both the BCP and GCP, indicating that this pedestal is likely ballooning-limited. %

\subsection{STAR}

The Spherical Tokamak Advanced Reactor (STAR) is an $A=2$, $R_0= 4$m compact high-field tokamak \cite{Menard2023_IAEA} targeting 100-500 MWe net electric power. Here, $A = R_0/a$ is the aspect-ratio for minor radius and major radius $R_0$. In \Cref{fig:phasebcpgcp}, the STAR device is shown to achieve high values of $\beta_{\theta, \mathrm{ped}}$ while having a narrower pedestal. Similar to NSTX discharge 132588, STAR relies on accessing KBM second-stability to obtain its steeper kinetic-ballooning-constrained pedestals, which shown in \Cref{fig:phasebcpgcp}, causes the GCP to give a steeper pedestal prediction than the GCP. A final $\beta_{\theta,\mathrm{ped}}$, $\Delta_{\mathrm{ped}}$ design point has not yet been determined. Notably, STAR achieves a relatively steeper pedestal measured approximately by $ \beta_{\theta, \mathrm{ped} }/ \Delta_{\mathrm{ped} }$ due to its low magnetic shear values across most of the pedestal, causing the KBM and ideal-ballooning-mode to be second stable.

\section{Aspect-Ratio and Shaping Scan} \label{sec:4}

In this section, we describe briefly aspect-ratio and shaping scans on ELMy NSTX discharge 132543 \cite{Maingi2015}. We choose a Luce parameterization for the last-closed-flux-surface shape \cite{Luce2013}. We define the flux surface elongation $\kappa$ and triangularity $\delta$ as the average $\langle \ldots \rangle_L$ of the Luce parameters $\kappa = \langle \kappa \rangle_L, \;\; \delta = \langle \delta \rangle_L$. The gk\_ped tools can perform aspect-ratio and shaping scans and find the BCP and GCP scaling expressions from the resulting new equilibria. The shaping and aspect-ratio scaling results for a different NSTX equilibrium are detailed in \cite{Parisi_arxiv2023_2}.

\begin{figure}
    \centering
    \begin{subfigure}[t]{0.79\textwidth}
        \centering
        \includegraphics[width=\textwidth]{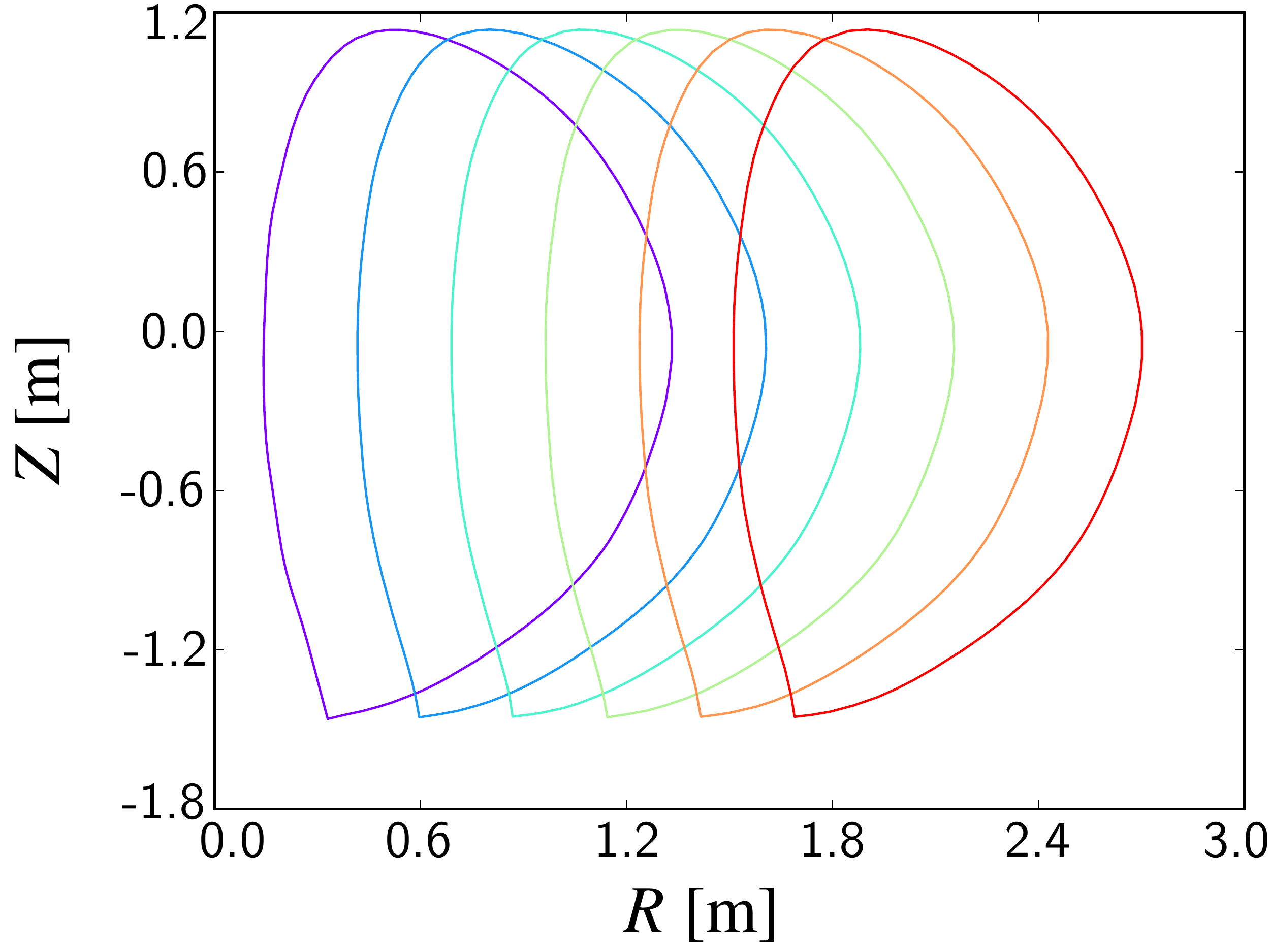}
        \caption{Last-closed-flux-surfaces across aspect-ratio.}
    \end{subfigure}
    ~
    \begin{subfigure}[t]{\textwidth}
        \centering
        \includegraphics[width=\textwidth]{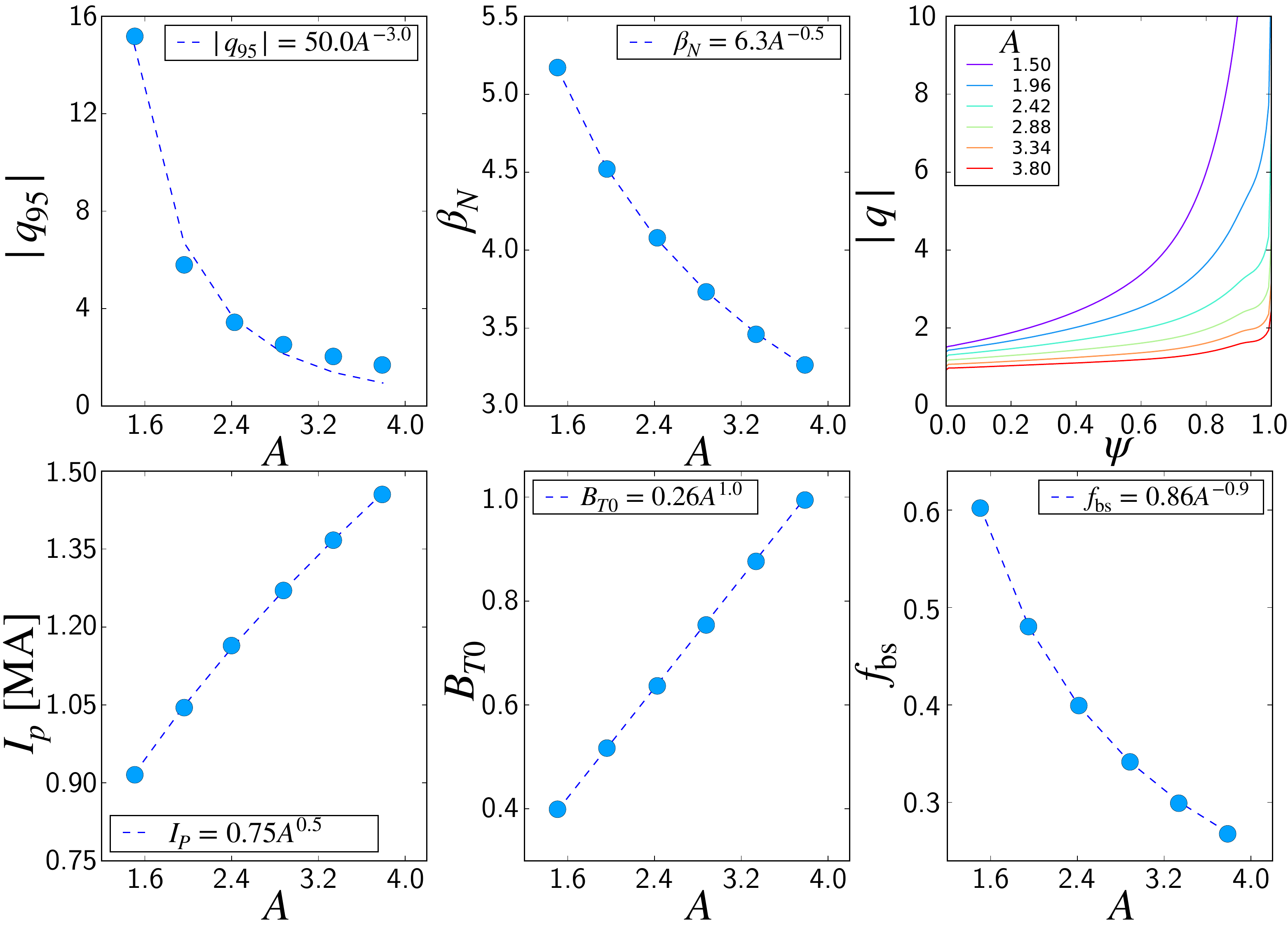}
        \caption{Top row, left to right: $q_{95}$, $\beta_N$, $q$. Bottom row: $I_p$, $B_{T0}$, $f_{\mathrm{BS}}$.}
    \end{subfigure}
    \caption{Equilibrium quantities for aspect-ratio scan in NSTX 132543.}
    \label{fig:AReq}
\end{figure}

For aspect-ratio scans, we use EFIT-AI \cite{Lao2022} to construct new equilibria for a range of aspect-ratios keeping the minor radius $a$ constant but allowing the major radius $R_0$ to vary. In \Cref{fig:AReq}(a), we plot the last-closed-flux-surfaces for $A \in [1.5, 1.96, 2.42, 2.88, 3.34, 3.8]$ -- for each aspect-ratio, a new magnetic coilset is generated by modifying the aspect-ratio of a basecase NSTX coilset, which is used to generate the Green's functions needed for free-boundary equilibrium reconstruction by the EFUND code \cite{Appel2006}. The boundary points are fixed and the equilibrium parameters are rescaled as follows $\beta_N \sim 1/\sqrt{R_0}, \;\; B_T \sim R_0, \;\; I_p \sim \sqrt{R_0}$. This gives $q_{95} \sim R_0^{-3}$ and bootstrap fraction, $f_{\mathrm{bs}} \sim R_0^{-0.9}$. In \Cref{fig:AReq}(b), we show how these quantities vary with aspect-ratio. For the highest major radius $R_0 \simeq 2.4$m, the flux-function $R_0 B_{T0} \simeq 2.4$[Tm], which is comparable to or lower than JET \cite{Matthews2011}, SPARC \cite{Creely2020}, and DIII-D \cite{Luxon2002}. Thus, our high-aspect-ratio equilibria respect reasonable engineering requirements for $R_0 B_{T0}$.

For triangularity scans, we fit the last-closed-flux-surface using a Luce parameterization and rescale $\delta = (\delta_{\mathrm{upper} } + \delta_{\mathrm{lower} })/2$ by a scalar factor. When varying triangularity, we keep all other plasma parameters constant. For elongation scans, we choose to vary the plasma current in order to keep $\beta_N$ constant. Given $\beta_T \sim (1 + \kappa^2) \beta_N^2$, in order to keep $\beta_N$ constant we vary plasma current as $I_p \sim 1 + \kappa^2$ at fixed $A$, $a$, $B_T$ \cite{Menard2016}. %

Once the equilibria with different aspect-ratio and shaping are generated, we evaluate the Gyrokinetic Critical Pedestal for each equilibrium.

\subsection{Least-Squares Fit}

 \begin{figure}[tb]
    \centering
    \includegraphics[width=\textwidth]{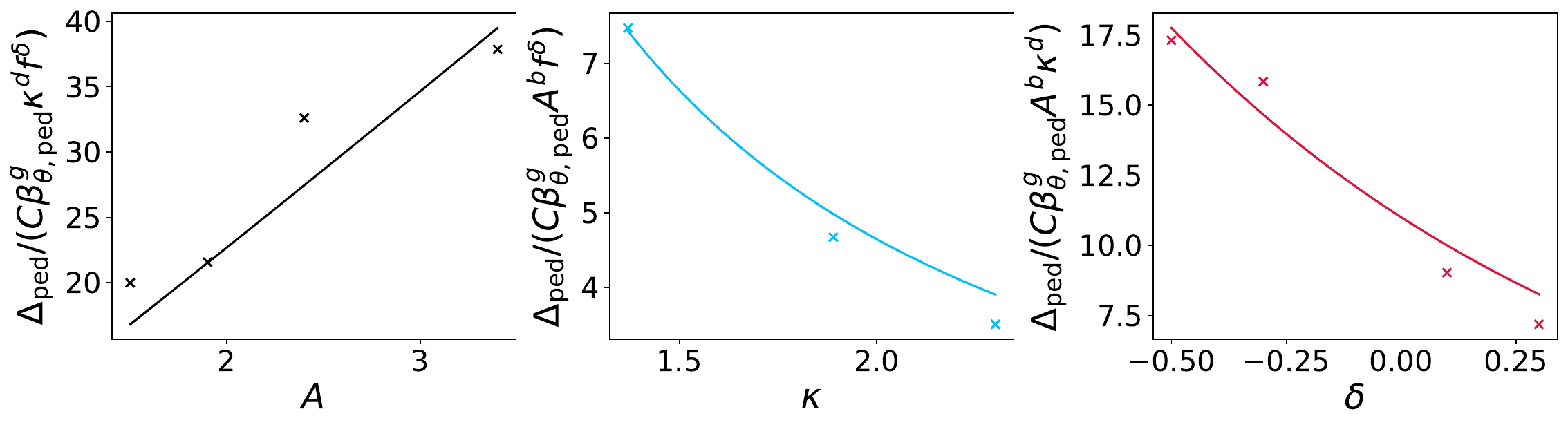}
        \caption{Fitting of $\Delta_{\mathrm{ped}}$ in \Cref{eq:bestfitform} to 11 shaping variations of NSTX discharge 132543, where the dependence on $A$, $\kappa$, and $\delta$ is shown in each subplot. For simplicity, we only show points where the x-axis parameter has changed from the nominal value.}
    \label{fig:fitting_plots}
\end{figure}

Based on our shaping and aspect-ratio scans, we perform a least-squares fit for the pedestal width of the form
\begin{equation}
\begin{aligned}
& \Delta_{\mathrm{ped}} = C A^{b} \kappa^{d} f^{\delta} \left( \beta_{\theta, \mathrm{ped}} \right)^g,
\end{aligned}
\label{eq:bestfitform}
\end{equation}
using the GCP from 11 shaping scans for NSTX 132543. In this paper, we fit the GCP width-scaling for only the wide GCP branch, finding
\begin{equation}
    \Delta_{\mathrm{ped}} = 0.92 A^{1.04} \kappa^{-1.24} 0.38^{\delta} \beta_{\theta,\mathrm{ped}}^{1.05},
\label{eq:delta_ped}
\end{equation}
with functional dependencies similar to fits performed in \cite{Parisi_arxiv2023_2} for a different NSTX equilibrium. The fitting parameters in \Cref{eq:delta_ped} have the values and standard deviation uncertainty: $C = 0.92 \pm 0.16$, $b = 1.04 \pm 0.16$, $d = -1.24 \pm 0.26$, $f = 0.38 \pm 0.04$, $g = 1.05 \pm 0.12$. In \Cref{fig:fitting_plots}, we plot the normalized width versus fitting and shaping parameters $A$, $\kappa$, and $\delta$. The $R^2$ value for all fitting parameters is $R^2 = 0.94$, indicating a relatively good fit. Removing any one of $A$, $\kappa$, $\delta$ reduced the $R^2$ value significantly, indicating the importance of all three of these parameters. This analysis can be improved in future work by: (1) increasing sample size, (2) performing fits on multi-discharge and multi-device experimental database, (3) a comprehensive assessment of more shaping and plasma parameters to find the most important parameters determining $\Delta_{\mathrm{ped}}$.

\section{Turbulent Transport Near The Gyrokinetic Critical Pedestal} \label{sec:transport}

\begin{table}
\centering
\caption{Aspect-ratio scan parameter values for NSTX 139047 used in \Cref{sec:transport}.} %
  \begin{tabular}{| c || ccc |}
    \hline
    Name & $A = R_0/a$ & $B_T$ [T] & $I_p$ [MA] \\ \hline
    $A_1$ & $1.6$ & $0.45$ & $0.93$ \\ \hline
    $A_2$ & $2.0$ & $0.55$ & $1.0$ \\ \hline
    $A_3$ & $2.5$ & $0.68$ & $1.1$ \\ \hline
    $A_4$ & $2.9$ & $0.81$ & $1.2$ \\ \hline
  \end{tabular}
\label{tab:tab_aspect}
\end{table}

In this section, we relate the pedestal width scaling to turbulent transport. We also use transport ratios from linear gyrokinetic simulations to study turbulent transport in the vicinity the GCP first and second stable branches. In these regions, the heat to particle transport ratios from KBM stability can vary significantly compared with transport in strongly-driven KBM regions. Given that experimentally the pedestal is often close to the GCP and far from the strongly-driven KBM regions, the changing transport properties of KBM and other modes has implications for the evolution of density and temperature profiles. In this section, we use equilibria based on NSTX discharge 139047 that have a range of aspect-ratio values, detailed in \Cref{tab:tab_aspect}. Unless mentioned otherwise, the nominal case used is NSTX discharge 139047 with a slightly increased aspect-ratio $A = 2.0$, referred to as $A_2$.

\subsection{Gyrokinetic Stability and Transport}

\begin{figure}[tb]
    \centering
    \includegraphics[width=\textwidth]{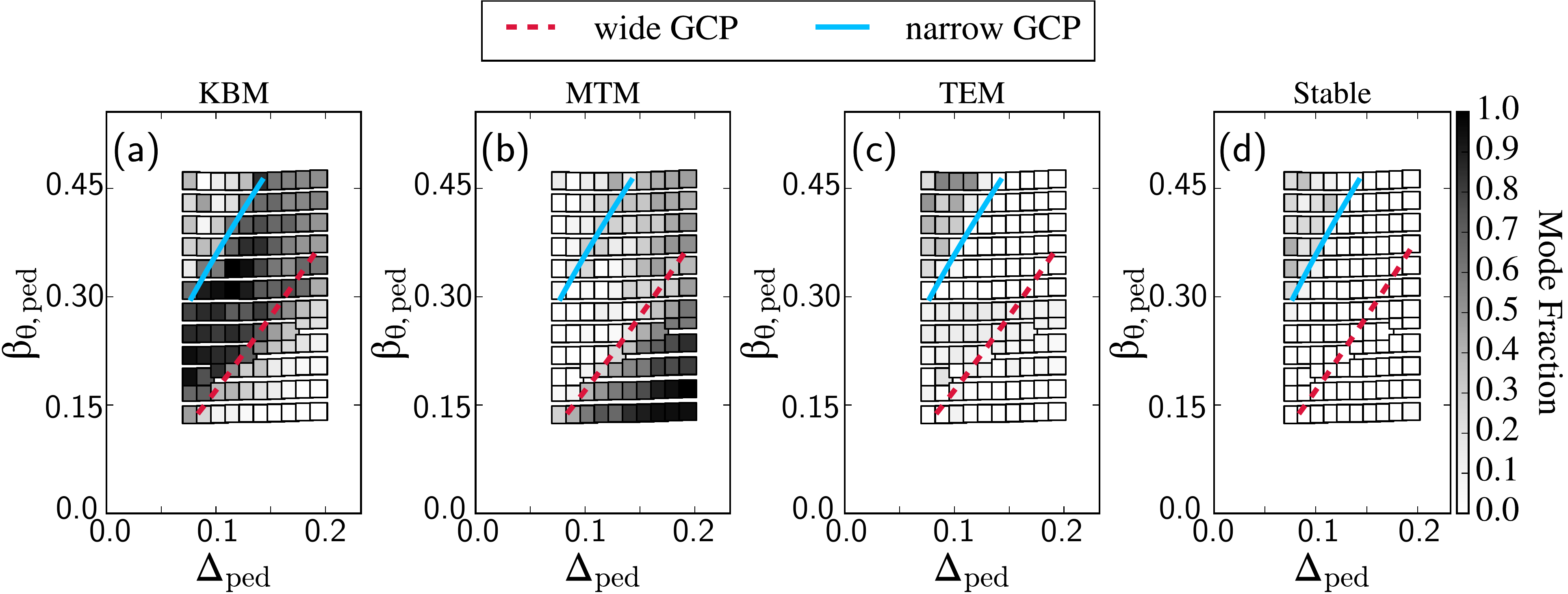}
        \caption{Fraction of gyrokinetic mode types across the pedestal half-width for fixed $n_{e,\mathrm{ped}}$ in NSTX 139047 $A_2$. The mode fraction is the fraction of all linear gyrokinetic simulations across the pedestal half-width and binormal wavenumbers (here $k_y \rho_i \in [0.06, 0.12, 0.18]$) where the fastest growing mode is classified as KBM, MTM, etc.}
    \label{fig:modetypes_AR2p0}
\end{figure}

Linear gyrokinetic simulations provide information about dominant mode type around the GCP. In \Cref{fig:modetypes_AR2p0}, we plot the fraction of different mode types across the pedestal half-width for binormal wavenumbers $k_y \rho_i \in [0.06, 0.12, 0.18]$ for NSTX discharge 139047 $A_2$. In the GCP unstable region, KBM dominates, with mode fractions lying in $0.5 -1.0$. At pressures below the wide GCP branch, MTM is ubiquitous and for pressures above the narrow GCP branch, TEM is the most common mode. We expect TEM instability generally for higher $k_y \rho_i$ values than included in these simulations, which explains the relatively low TEM fraction at pressures above the narrow GCP in \Cref{fig:modetypes_AR2p0}(d).

In addition to mode type, linear gyrokinetic simulations give flux and diffusivity ratios. The gyroBohm-normalized heat and particle fluxes through a flux surface are
\begin{equation}
\begin{aligned}
& q_s = \langle \int \frac{ m_s v^2}{2} h_s \mathbf{v_{\chi}} \cdot \frac{ \nabla \psi}{|\nabla \psi|} d^3v \rangle_{\psi} / q_{\mathrm{ gB}}  = \chi_s \frac{a}{L_{T_s}} + \frac{3}{2} \Gamma_s , \\ 
& \Gamma_s = \langle \int h_s \mathbf{v_{\chi}} \cdot \frac{ \nabla \psi}{|\nabla \psi|} d^3v \rangle_{\psi}/ \Gamma_{\mathrm{gB}} = D_s \frac{a}{L_{n_s}},
\end{aligned}
\label{eq:q_and_gamma}
\end{equation} 
where $\langle \ldots \rangle_{\psi}$ is a flux-surface average, $m_s$ is the particle mass, $q_s$ and $\Gamma_s$ are the gyroBohm normalized heat and particle fluxes for a species $s$, $\chi_s$ and $D_s$ are the normalized heat and particle diffusivities, and $q_{\mathrm{gB} } = \rho_{*r}^2 n_r T_r \bar{c}, \; \Gamma_{gB} = \rho_{*r}^2 n_r \bar{c}$ where $r$ subscripts refer to a reference species, $\bar{c} = \sqrt{T_e / m_D}$ is the sound speed, and $\rho_{*r} = \rho_r / a$ where $\rho_r$ is the gyroradius. %
The ratio of heat to particle flux is
\begin{equation}
\frac{q_s}{\Gamma_s}=\eta_{s} \frac{\chi_s}{D_s} + \frac{3}{2},
\label{eq:eta_flux_ratio}
\end{equation}
where $\eta_s = \nabla \ln T_{s}/\nabla \ln n_s$. Thus, changing $\eta_s$ by varying pedestal height via density or temperature can affect strongly the heat and particle flux ratios, even if $\chi_s / D_s$ is constant. %

\begin{figure*}[tb]
    \centering
    \begin{subfigure}[t]{0.31\textwidth}
        \centering			
        \includegraphics[width=\textwidth]{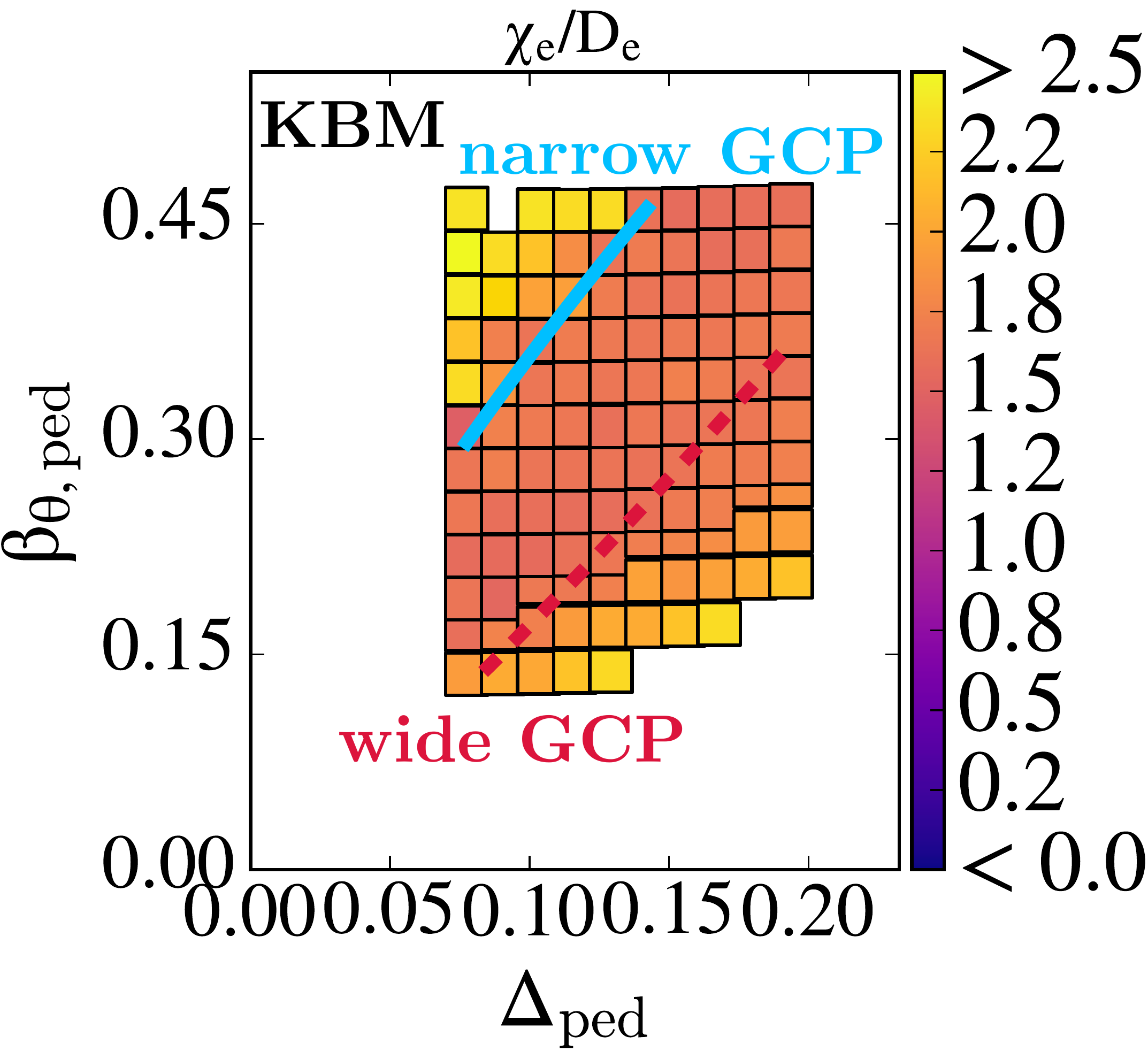}
        \caption{KBM $\chi_e / D_e$.}
    \end{subfigure}
    ~
    \begin{subfigure}[t]{0.31\textwidth}
        \centering
        \includegraphics[width=\textwidth]{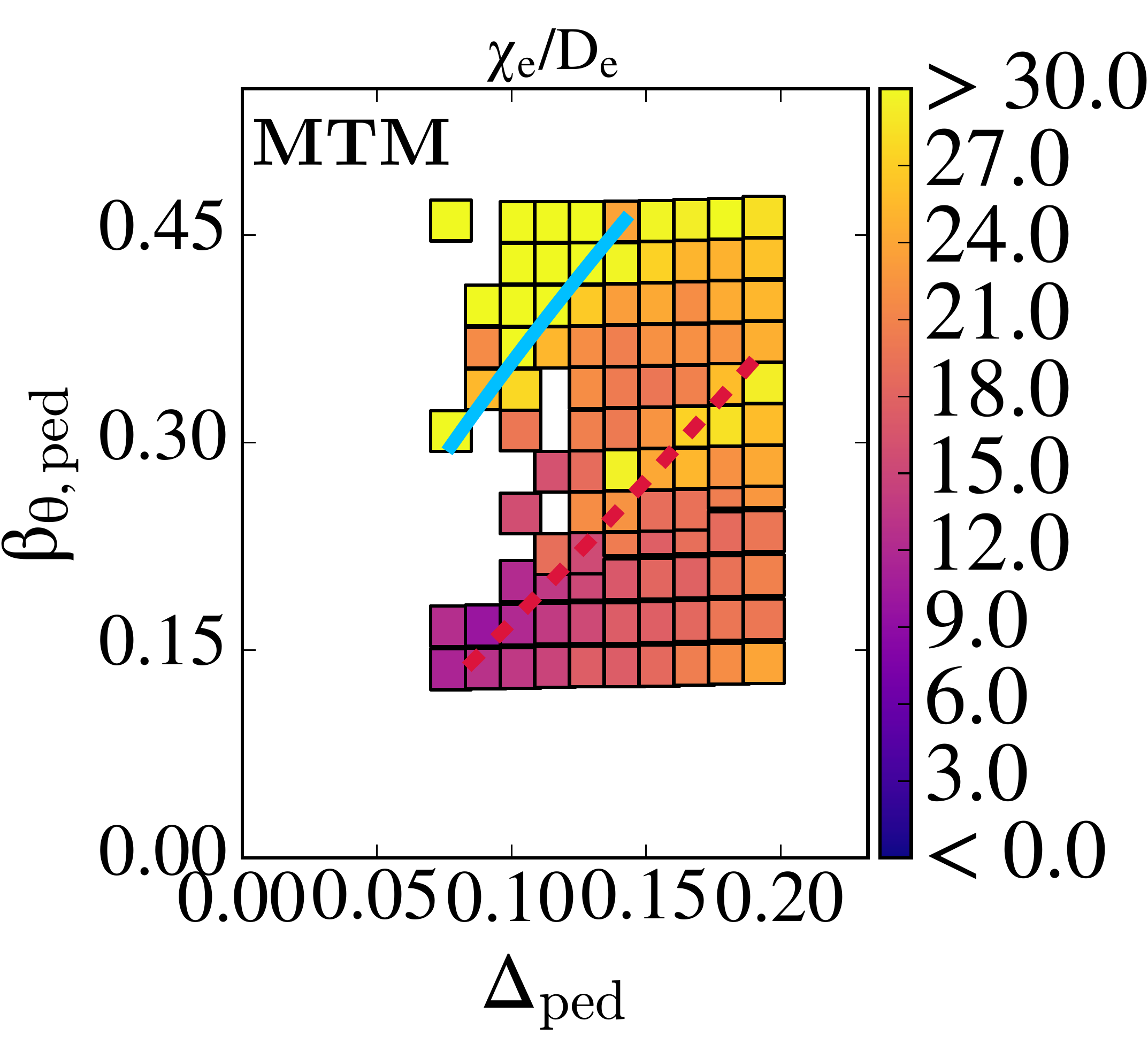}
        \caption{MTM $\chi_e / D_e$.}
    \end{subfigure}
    ~
    \begin{subfigure}[t]{0.31\textwidth}
        \centering
        \includegraphics[width=\textwidth]{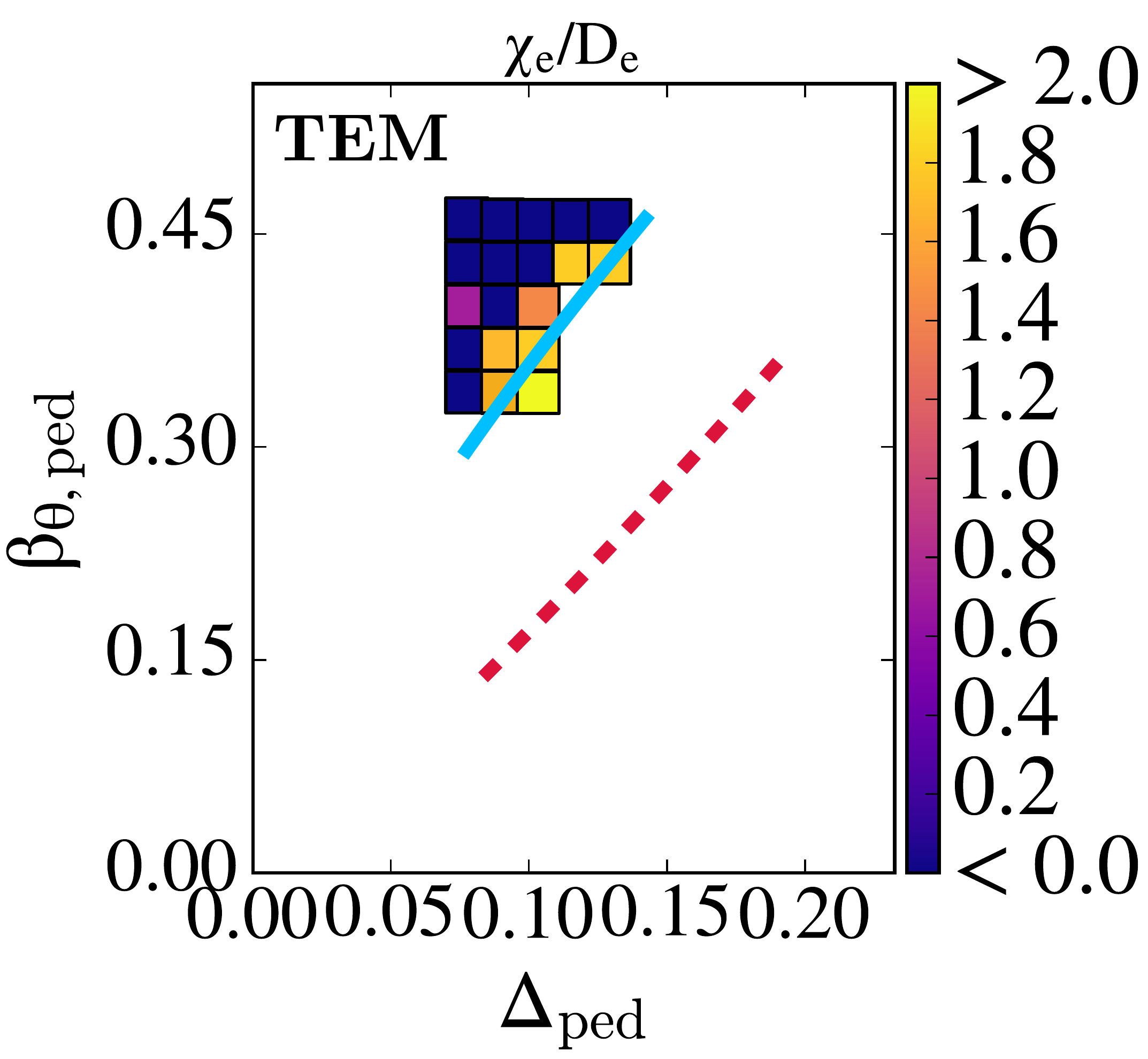}
        \caption{TEM $\chi_e / D_e$.}
    \end{subfigure}
    \begin{subfigure}[t]{0.31\textwidth}
        \centering
        \includegraphics[width=\textwidth]{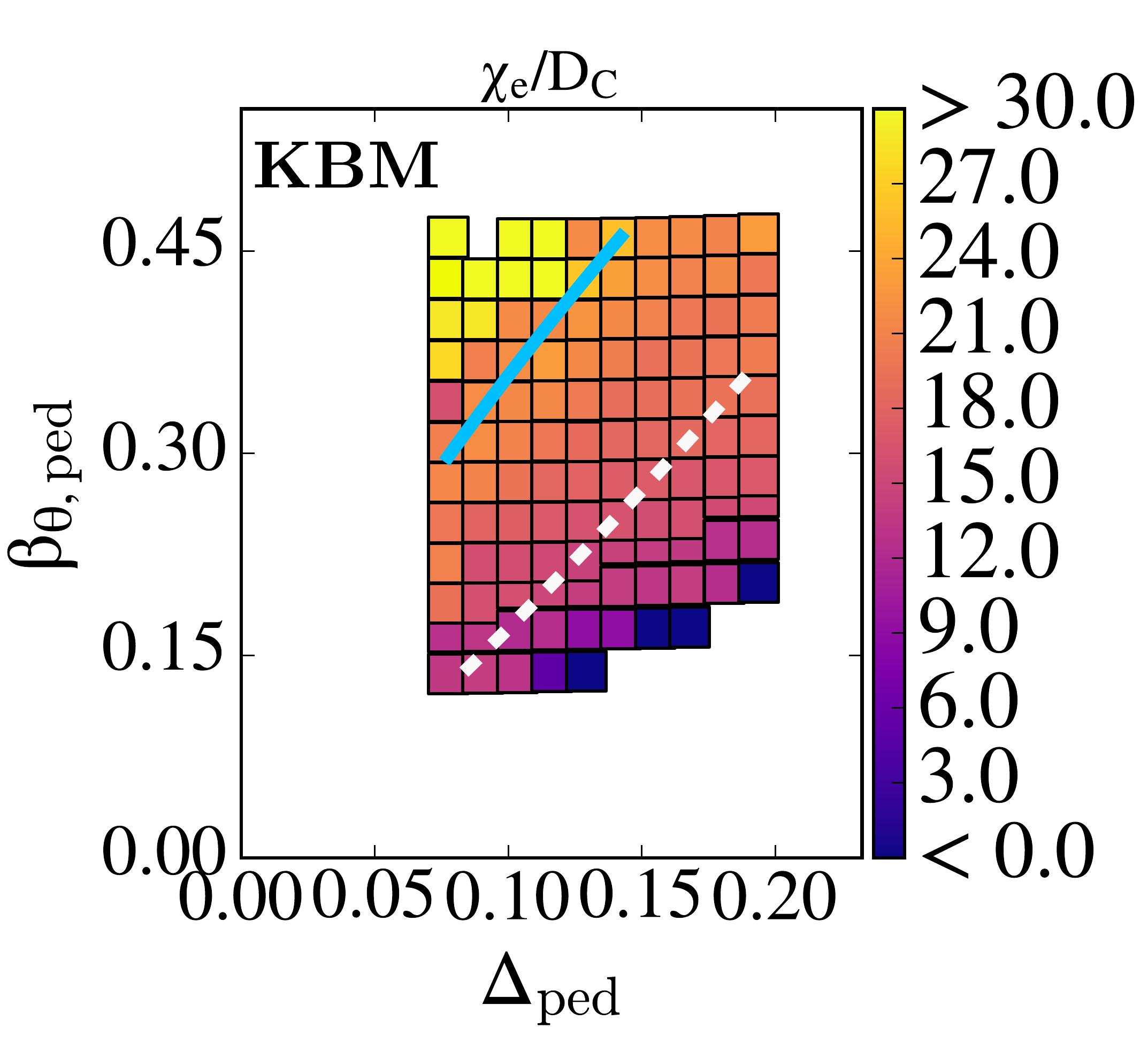}
        \caption{KBM $\chi_e / D_C$.}
    \end{subfigure}
    ~
    \begin{subfigure}[t]{0.31\textwidth}
        \centering
        \includegraphics[width=\textwidth]{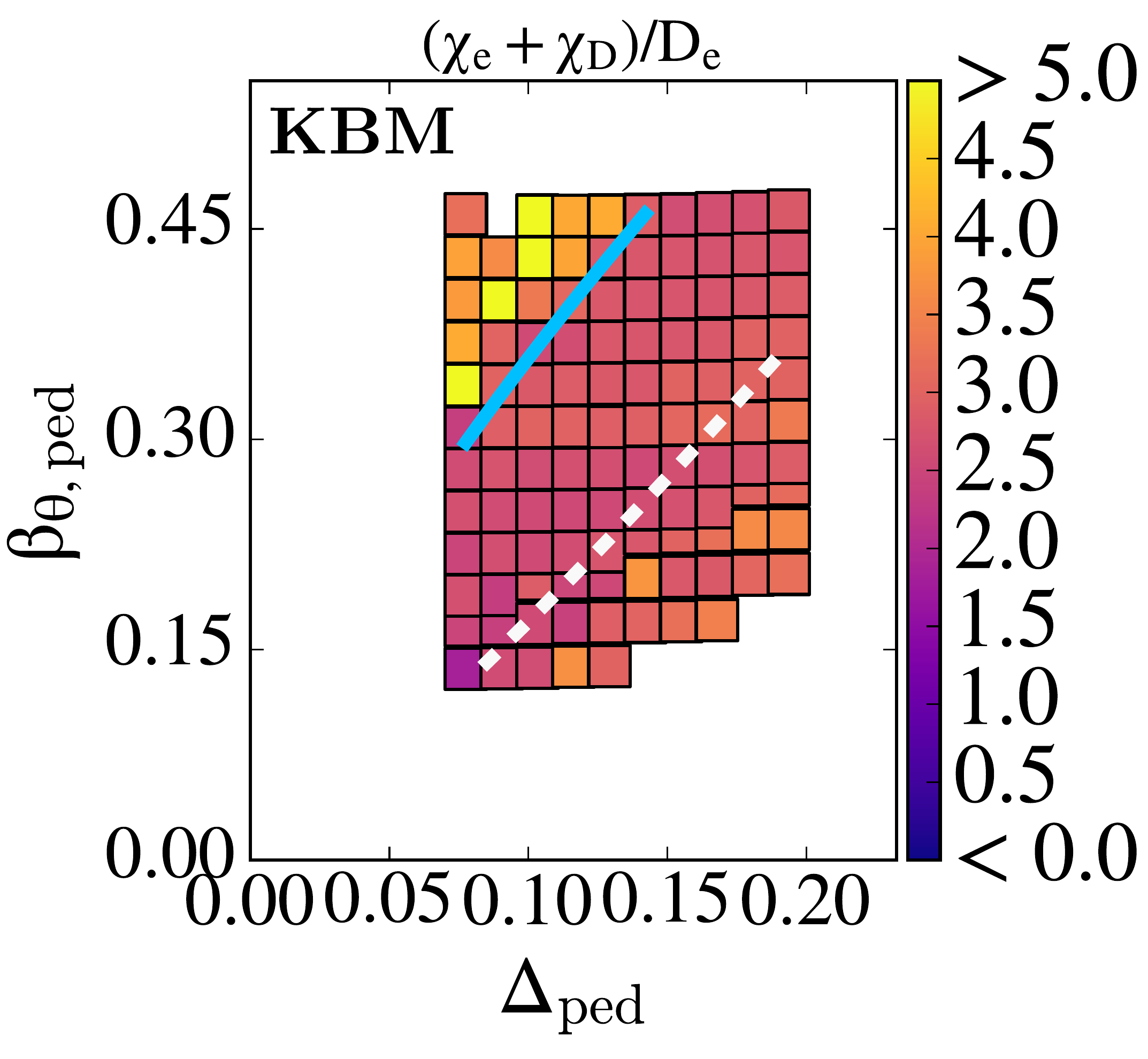}
        \caption{KBM $(\chi_e + \chi_D)/ D_e$.}
    \end{subfigure}
    \caption{Top row: electron diffusivities $D_e/\chi_e$ for KBM, MTM, and TEM for NSTX 139047 $A_2$. Bottom row: KBM $\chi_e / D_C$ and $(\chi_e + \chi_D)/ D_e$.}
    \label{fig:modetypes_test1p5}
\end{figure*}

We find the ratio $\chi_e / D_e$ varies significantly around the GCP. In the top row of \Cref{fig:modetypes_test1p5}, we plot $\chi_e / D_e$ versus $\beta_{\theta,\mathrm{ped}}$ and $\Delta_{\mathrm{ped}}$ for KBM, MTM, and TEM instability in NSTX discharge 139047 $A_2$. In strongly driven KBM regions of \Cref{fig:modetypes_test1p5}(a), $\chi_e / D_e \simeq 1.5$. However, as the KBM is stabilized near the narrow and wide GCP branches, its transport coefficients change, often satisfying $\chi_e / D_e \simeq 2 - 5$. For MTM instability in \Cref{fig:modetypes_test1p5}(b) $\chi_e / D_e \simeq 10 - 50$ and for TEM in \Cref{fig:modetypes_test1p5}(c) $|\chi_e / D_e| \simeq 0- 2$. While nonlinear simulations \cite{Pueschel2008} are required for accurate flux ratios, the strong variation of $\chi_e/D_e$ close to marginal stability presents challenges to reduced transport-based pedestal models \cite{Guttenfelder2021, Hatch2023}.

Around the GCP, $\chi_e / D_C$ is much larger than reported in the literature \cite{Kotschenreuther2019} where it was reported $\chi_e / D_C \simeq 3/2$. Here, $D_C$ is the Carbon-12 particle diffusivity. Plotted in \Cref{fig:modetypes_test1p5}(d), along the wide GCP $\chi_e / D_C \simeq 15-20$, indicating that KBM produces relatively weak impurity transport in the inter-ELM period. We use the ratio $(\chi_e + \chi_D)/ D_e$ \cite{Kotschenreuther2019} to distinguish between KBM and TEM, where $\chi_D$ is the heat diffusivity for the main ion deuterium. For KBM along the GCP we find $(\chi_e + \chi_D)/ D_e \simeq 3$, increasing substantially near marginal KBM stability to $(\chi_e + \chi_D)/ D_e \simeq 10$. 

The transport coefficients in different GCP regions are summarized later in \Cref{tab:tab_transp_coeffs} and are discussed more in \Cref{sec:transp_marginal}.

\begin{figure*}[tb]
    \centering
    \begin{subfigure}[t]{0.4\textwidth}
        \centering
        \includegraphics[width=\textwidth]{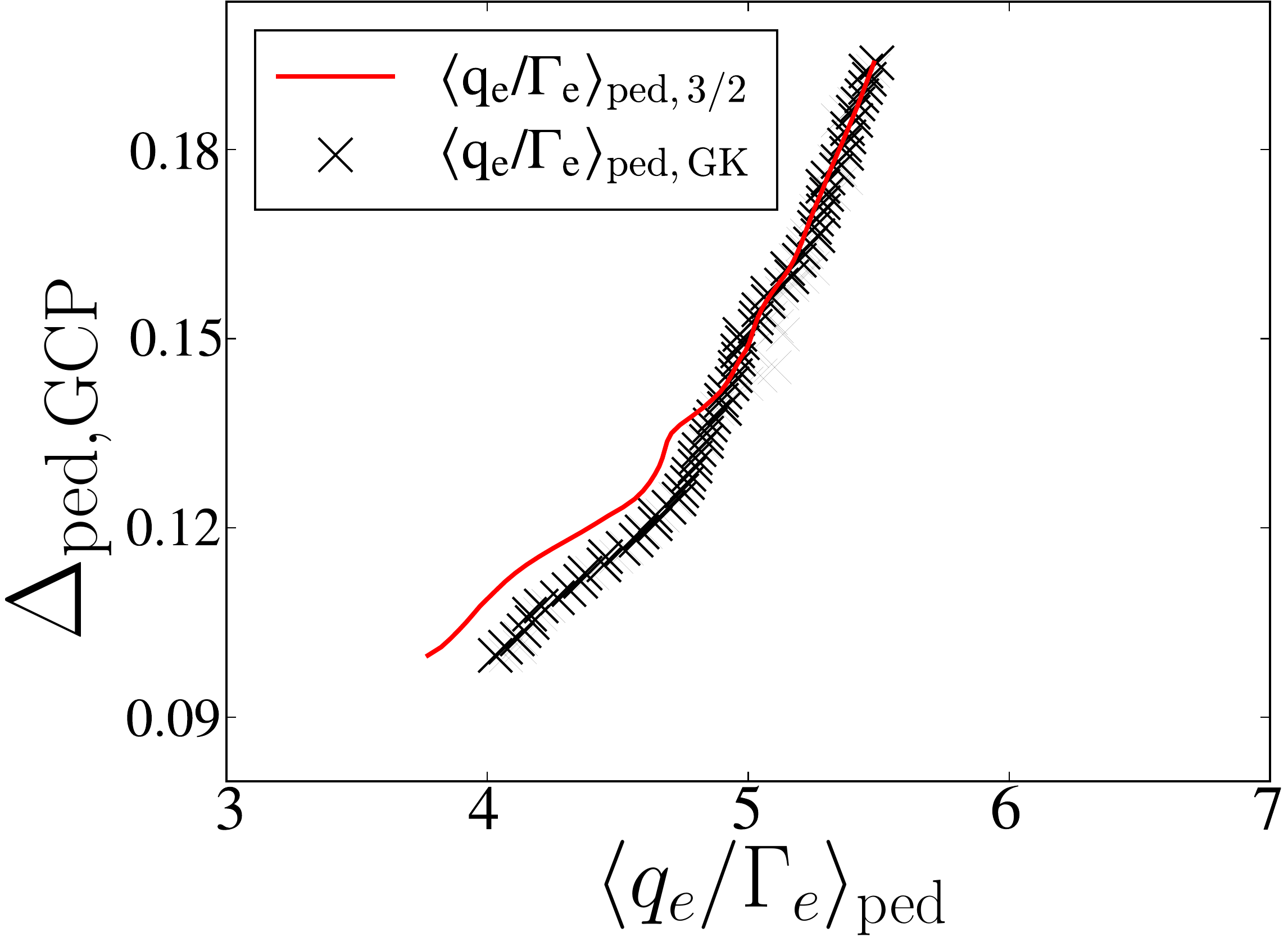}
        \caption{Width versus $\langle q_e/\Gamma_e \rangle_{\mathrm{ped} }$ along wide GCP branch at fixed $n_{e,\mathrm{ped}}$ for transport model $\langle q_e/\Gamma_e \rangle_{\mathrm{ped,3/2} }$ and simulations $\langle q_e/\Gamma_e \rangle_{\mathrm{ped, GK} }$.}
    \end{subfigure}
    ~
    \begin{subfigure}[t]{0.5\textwidth}
        \centering
        \includegraphics[width=\textwidth]{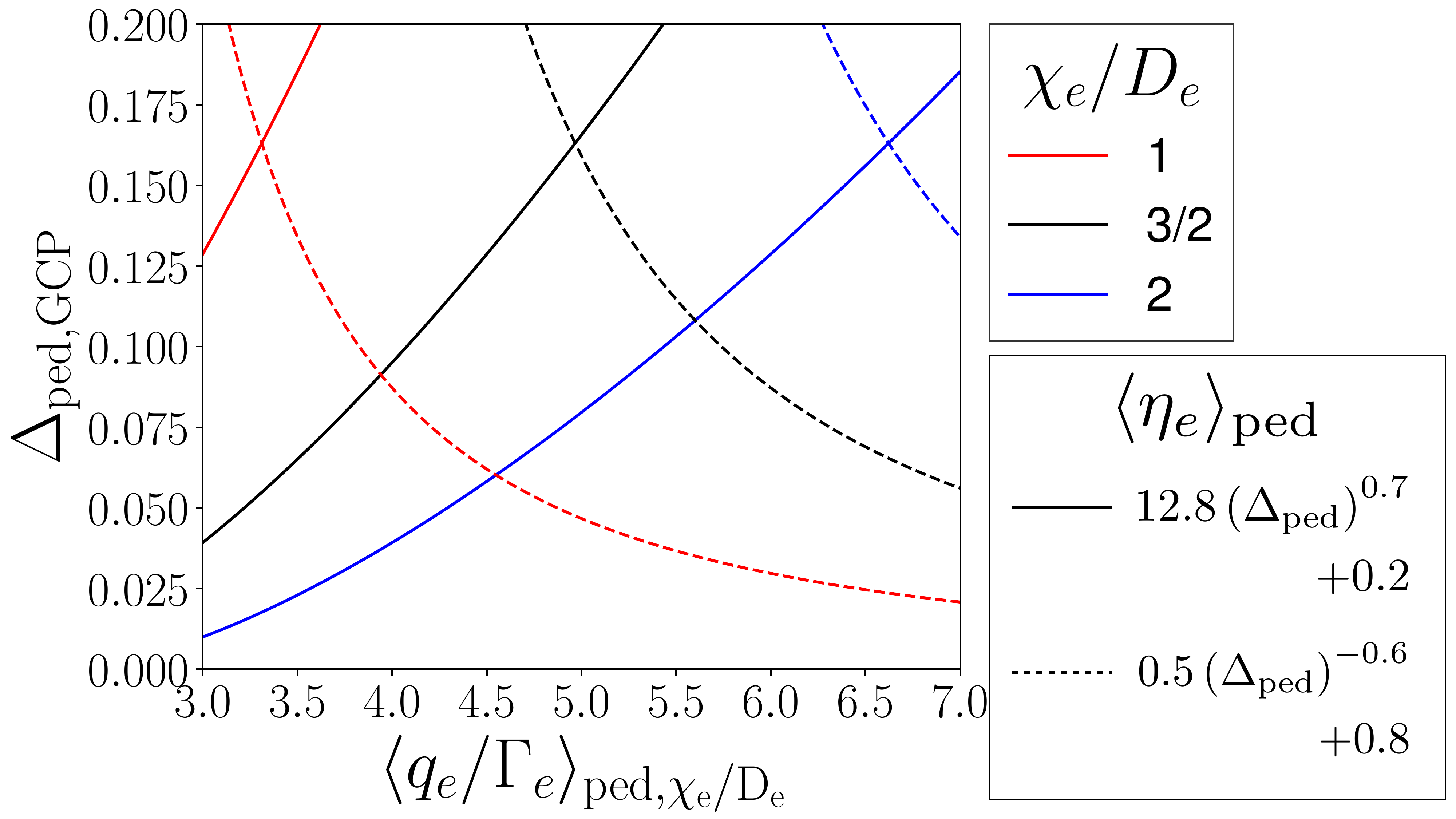}
        \caption{Width prediction using two different fits of $\langle \eta_e \rangle_{\mathrm{ped}} $ for three $\chi_e/ D_e \in [1,1.5,2.0]$ values for the KBM.}
    \end{subfigure}
    \caption{Pedestal width along the GCP versus $\langle q_e/\Gamma_e \rangle_{\mathrm{ped} }$ from gyrokinetic simulations and a transport model in \Cref{eq:width_scaling_general} for NSTX 139047 $A_2$. In (a), we use the exact values of $\langle \eta_e \rangle_{\mathrm{ped} } $ to calculate $\langle q_e/\Gamma_e \rangle_{\mathrm{ped,3/2} }$. In (b), a best fit is used for $\langle q_e/\Gamma_e \rangle_{\mathrm{ped,\chi_e/ D_e} }$.}
    \label{fig:etaescaling_heat_single}
\end{figure*}

\subsection{Width-Transport Scaling}

In this section, we take initial steps in relating pedestal width and transport. While linear stability threshold models \cite{Snyder2009,Parisi2023_ARXIV} provide width-height scalings, they omit the sources and transport required to sustain pedestal profiles. Here, we find the dependence of pedestal width on the transport ratio $q_e / \Gamma_e$. To simplify analysis, we study quantities averaged over the pedestal half-width,
\begin{equation}
\left\langle \frac{q_e}{\Gamma_e} \right\rangle_{\mathrm{ped}} = \int_{\psi_{1/4}}^{\psi_{3/4 } } \frac{q_e}{\Gamma_e} \; d \psi / \int_{\psi_{1/4}}^{\psi_{3/4 } } d \psi = (2/\Delta_{\mathrm{ped}}) \int_{\psi_{1/4}}^{\psi_{3/4 } } \frac{q_e}{\Gamma_e} \; d \psi,
\label{eq:ped_half_average}
\end{equation}
where $\psi_{1/4} = \psi_{\mathrm{mid} } - \Delta_{\mathrm{ped}}/4$ and $\psi_{3/4} = \psi_{\mathrm{mid} } + \Delta_{\mathrm{ped}}/4$. %
Along the GCP, if KBM dominates electron heat and particle transport and $\chi_{e,\mathrm{KBM} }/D_{e,\mathrm{KBM} } = 3/2$ is constant, radially averaging \Cref{eq:eta_flux_ratio} gives
\begin{equation}
\left\langle \frac{q_s}{\Gamma_s} \right\rangle_{\mathrm{ped, 3/2} }= \frac{3}{2} \left( \langle \eta_e \rangle_{\mathrm{ped} } + 1 \right),
\label{eq:eta_flux_ratio_av}
\end{equation}
where the $3/2$ subscript in $\langle q_e/\Gamma_e \rangle_{\mathrm{ped,3/2}}$ indicates that we assumed $\chi_{e}/D_{e } = 3/2$. In \Cref{fig:etaescaling_heat_single}(a), we plot $\langle q_e/\Gamma_e \rangle_{\mathrm{ped,3/2} }$ along the GCP wide branch using \Cref{eq:eta_flux_ratio_av} for NSTX 139047 $A_2$ and also plot $\langle q_e/\Gamma_e \rangle_{\mathrm{ped,GK} }$ using data from KBMs in gyrokinetic simulations,
\begin{equation}
\left\langle \frac{q_s}{\Gamma_s} \right\rangle_{\mathrm{ped, GK}}  = \frac{ \sum_{k_y} \int_{\psi_{1/4}}^{\psi_{3/4 } } {q_e/\Gamma_e}_{\mathrm{KBM} } (\psi, k_y)  \; d \psi} {\sum_{k_y} \int_{\psi_{1/4}}^{\psi_{3/4 } } d \psi},
\end{equation}
where $\sum_{k_y}$ is a sum over all $k_y$ wavenumbers included in the simulation and ${q_e/\Gamma_e}_{\mathrm{KBM} }$ is evaluated only if the fastest growing mode is a KBM. The close agreement between $\langle q_e/\Gamma_e \rangle_{\mathrm{ped,3/2}}$ and $\langle q_e/\Gamma_e \rangle_{\mathrm{ped,GK}}$ in \Cref{fig:etaescaling_heat_single}(a) demonstrates that $\chi_e/D_e = 3/2$ is an excellent approximation for the KBM along this particular wide branch GCP.
\begin{figure*}[tb]
    \centering
    \begin{subfigure}[t]{0.47\textwidth}
        \centering
        \includegraphics[width=\textwidth]{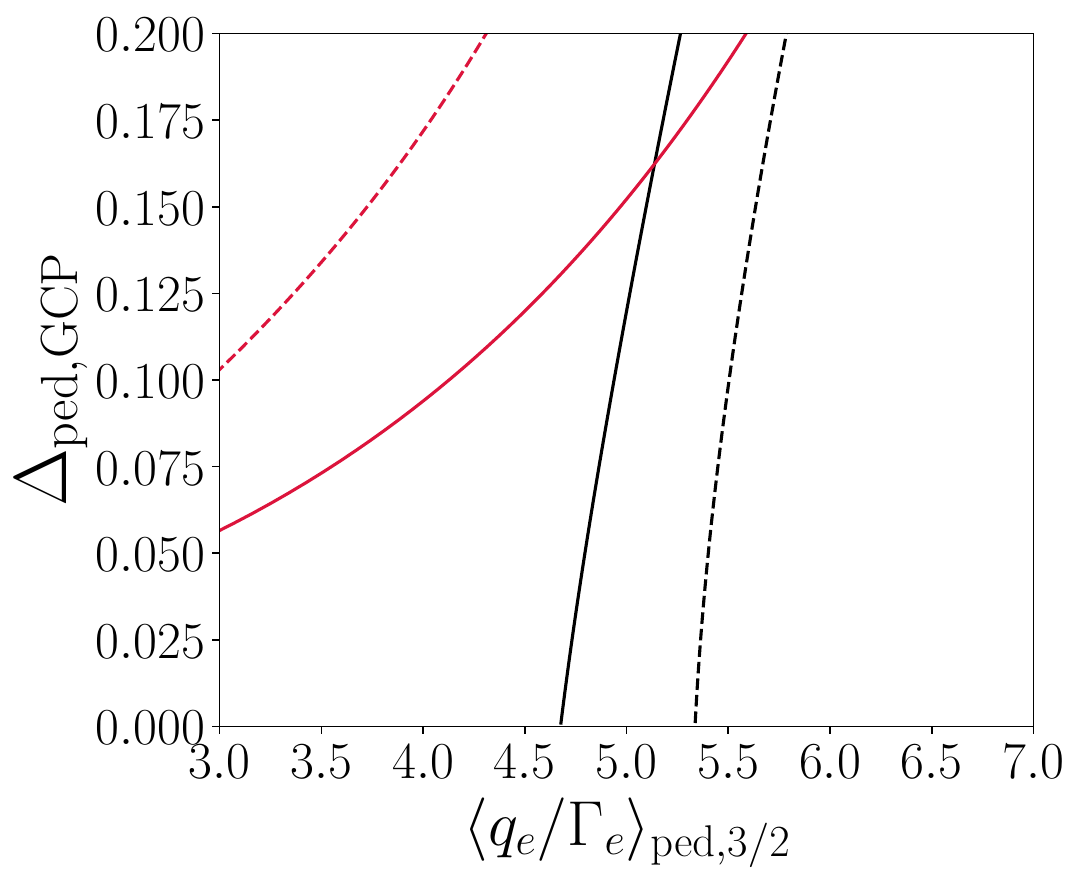}
        \caption{$\Delta_{\mathrm{ped} }$ versus $\langle q_e/\Gamma_e \rangle_{\mathrm{ped,3/2}}$.}
    \end{subfigure}
    ~
    \begin{subfigure}[t]{0.47\textwidth}
        \centering
        \includegraphics[width=\textwidth]{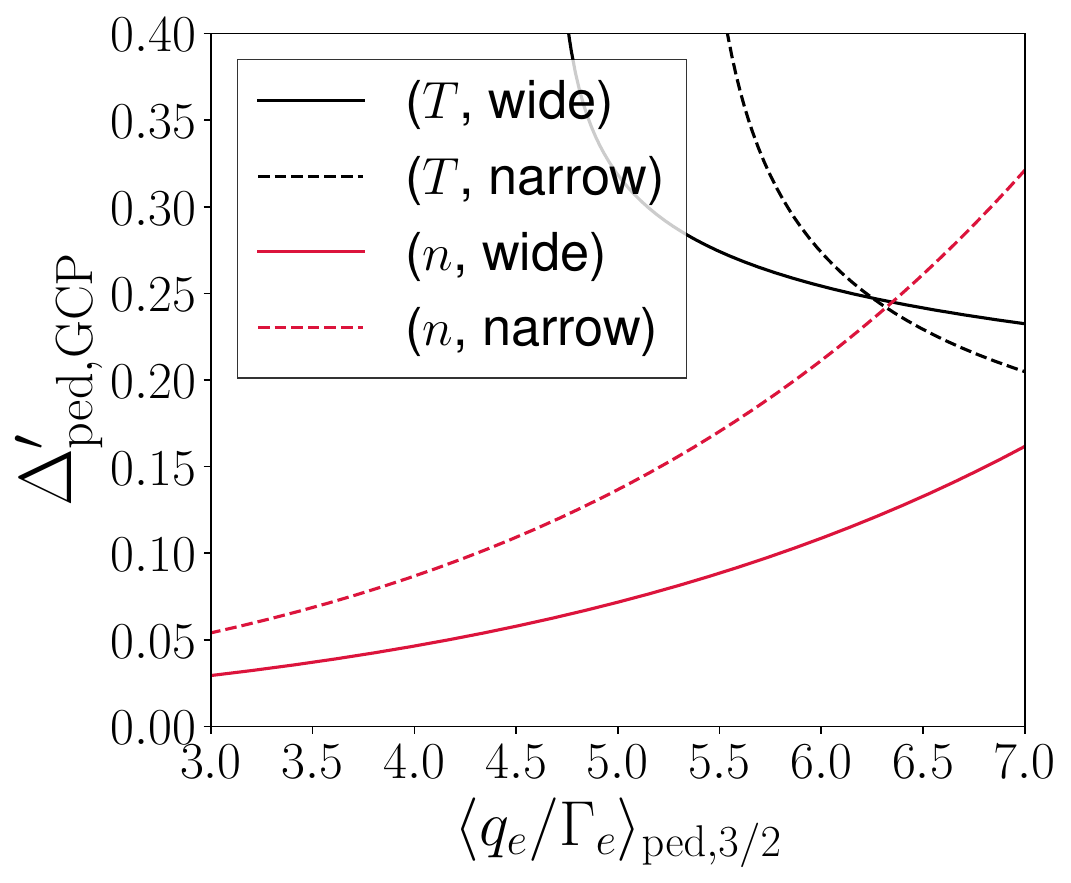}
        \caption{$\Delta_{\mathrm{ped} } '$ versus $\langle q_e/\Gamma_e \rangle_{\mathrm{ped,3/2}}$.}
    \end{subfigure}
    \caption{Pedestal width (\Cref{eq:width_scaling_general}) and its derivative $\Delta_{\mathrm{ped} } ' = d \Delta_{\mathrm{ped} }/ d \langle q_e/\Gamma_e \rangle_{\mathrm{ped,3/2}}$ as a function of $\langle q_e/\Gamma_e \rangle_{\mathrm{ped,3/2}}$ for NSTX 139047 $A_2$. We show four GCP branches: wide and narrow for fixed $n_{e,\mathrm{ped}}$ (red) and fixed $T_{e,\mathrm{ped}}$ (black).}
    \label{fig:etaescaling_heat}
\end{figure*}

We now obtain an expression for $\Delta_{\mathrm{ped,GCP}}$ as a function of $\langle q_e/\Gamma_e \rangle_{\mathrm{ped,3/2} }$ by relating $\langle \eta_e \rangle_{\mathrm{ped} }$ to $\Delta_{\mathrm{ped,GCP}}$ through $\beta_{\theta,\mathrm{ped,GCP}}$. It is useful to obtain $\Delta_{\mathrm{ped,GCP}}$ through $\eta_e$ because later we will examine how varying $\beta_{\theta, \mathrm{ped}}$ with constant $n_{\mathrm{ped} }$ or constant $T_{\mathrm{ped} }$ affects $\eta_e$ and therefore pedestal width. We write the dependence of $\langle \eta_e \rangle_{\mathrm{ped} } $ on $\beta_{\theta,\mathrm{ped,GCP}}$,
\begin{equation}
\langle \eta_e \rangle_{\mathrm{ped} } = a_1 \left( \beta_{\theta,\mathrm{ped,GCP}} \right)^{\alpha_1} + d = a_1 D^{\alpha_1} \left( \Delta_{\mathrm{ped,GCP}} \right)^{\alpha_1 \alpha_2} + d,
\label{eq:etae_relation}
\end{equation}
where we used the GCP relation $\beta_{\theta,\mathrm{ped,GCP}} = D \left( \Delta_{\mathrm{ped,GCP}} \right)^{\alpha_2}$. Substituting \Cref{eq:etae_relation} into \Cref{eq:eta_flux_ratio_av} gives $\Delta_{\mathrm{ped,GCP}}$ as a function of $\langle q_e/\Gamma_e \rangle_{\mathrm{ped,3/2} }$,
\begin{equation}
\Delta_{\mathrm{ped,GCP}} = \left( \frac{\frac{2}{3} \langle \frac{q_e}{\Gamma_e} \rangle_{\mathrm{ped,3/2} } - (1 + d) }{a_1 \left( D \right)^{\alpha_1}}  \right)^{1/(\alpha_1 \alpha_2)}.
\label{eq:width_scaling_general}
\end{equation}
For fixed $n_{\mathrm{ped}}$, we find $\alpha_1 = 0.55$, $\alpha_2 = 1.2$, $d = 0.165$, $a_1 = 4.33$, and $D = 2.47$, giving a wide GCP width-transport scaling
\begin{equation}
\Delta_{\mathrm{ped,wide \; GCP,fixed \; n}} \simeq 0.028 \left(\left\langle \frac{q_e}{\Gamma_e} \right\rangle_{\mathrm{ped,3/2} } - 1.7 \right)^{1.5}.
\label{eq:deltaped_KBMtransport}
\end{equation}
In \Cref{fig:etaescaling_heat_single}(b), in solid lines we plot $\Delta_{\mathrm{ped,GCP} }$ in \Cref{eq:width_scaling_general} for three values of $\chi_{e,\mathrm{KBM} }/D_{e,\mathrm{KBM} } \in [1,3/2,2]$ using the fitted form of $\langle \eta_e \rangle_{\mathrm{ped} }$ in \Cref{eq:etae_relation}: The pedestal width is very sensitive to the  $\chi_{e,\mathrm{KBM} }/D_{e,\mathrm{KBM} }$ value used in $\langle q_e / \Gamma_e \rangle_{\mathrm{ped}}$. Using $\chi_e/ D_e = 2$ we find $\Delta_{\mathrm{ped,GCP} } \simeq 0.025$ at $\langle q_e / \Gamma_e \rangle_{\mathrm{ped}} = 3.5$, but using $\chi_e/ D_e = 1$ we find $\Delta_{\mathrm{ped,GCP} } \simeq 0.2$. Therefore, a 2x decrease in $\chi_e/ D_e$ gives an 8x increase in $\Delta_{\mathrm{ped,GCP} }$.

In \Cref{fig:etaescaling_heat_single}(b), using dashed lines we also plot $\Delta_{\mathrm{ped,GCP} }$ for an $\langle \eta_e \rangle_{\mathrm{ped} }$ function that depends \textit{inversely} with $\Delta_{\mathrm{ped,GCP} }$: $\langle \eta_e \rangle_{\mathrm{ped} } = 0.5 \left( \Delta_{\mathrm{ped} } \right)^{-0.6} + 0.8$, but using the same width-height scaling relation. Such an inverse relation for $\Delta_{\mathrm{ped} }$ and $\langle \eta_e \rangle_{\mathrm{ped} }$ would result from varying $\beta_{\theta,\mathrm{ped} }$ at fixed $T_{e, \mathrm{ped} }$ and fixed $n_{e,\mathrm{sep}}$ (see \Cref{eq:general_etae}, \Cref{app:eta_rescaling}). The widest pedestals are given by the largest $\chi_{e,\mathrm{KBM} }/D_{e,\mathrm{KBM} }$ values rather than the smallest $\chi_{e,\mathrm{KBM} }/D_{e,\mathrm{KBM} }$, which happened when $\langle \eta_e \rangle_{\mathrm{ped} } $ increased with $\Delta_{\mathrm{ped,GCP} }$.

By varying pedestal height at fixed $T_{e,\mathrm{ped}}$ for NSTX discharge 139047 $A_2$, $\Delta_{\mathrm{ped,GCP} }$ is much more sensitive to $\langle \eta_e \rangle_{\mathrm{ped} }$, making $\Delta_{\mathrm{ped,GCP} }$ more sensitive to $\langle q_e / \Gamma_e \rangle_{\mathrm{ped}}$ at fixed $T_{e,\mathrm{ped}}$ than at fixed $n_{e,\mathrm{ped}}$. At fixed $T_{e,\mathrm{ped}}$, we find the width-transport scaling,
\begin{equation}
\Delta_{\mathrm{ped,wide \; GCP,fixed \; T}} \simeq 0.31 \left(\left\langle \frac{q_e}{\Gamma_e} \right\rangle_{\mathrm{ped,3/2} } - 4.7 \right)^{0.85}.
\label{eq:deltaped_KBMtransport_two}
\end{equation}
The stronger sensitivity of pedestal width to $\langle q_e / \Gamma_e \rangle_{\mathrm{ped}}$ at fixed  $T_{e,\mathrm{ped}}$ occurs because we keep $n_{\mathrm{ped} }/n_{\mathrm{sep}}$ fixed when increasing $\beta_{\mathrm{ped} }$, making $\eta_e$ only weakly dependent on $\beta_{\mathrm{ped} }$. In \Cref{fig:etaescaling_heat}(a) we plot $\Delta_{\mathrm{ped, GCP}}$ for fixed $n_{e,\mathrm{ped}}$ and fixed $T_{e,\mathrm{ped}}$ narrow and wide GCP branches using the fitted form of $\Delta_{\mathrm{ped, GCP}}$ in \Cref{eq:width_scaling_general} -- notably, the black curves corresponding to fixed $T_{e,\mathrm{ped}}$ narrow and wide GCP branches have relatively large derivatives, as shown in \Cref{fig:etaescaling_heat}(b). This suggests that for KBM-limited pedestals with fixed $T_{e,\mathrm{ped}}$, a small increase of $\langle q_e / \Gamma_e \rangle_{\mathrm{ped,3/2}}$ (e.g. decreased fueling) would increase $\Delta_{\mathrm{ped}}$ much more effectively than increasing $\langle q_e / \Gamma_e \rangle_{\mathrm{ped}}$ by increasing the heating. %

To summarize, \Cref{eq:deltaped_KBMtransport,eq:deltaped_KBMtransport_two} are the wide GCP pedestal width-transport expressions resulting from a simple transport model that assumes along the GCP, turbulent particle \textit{and} heat transport are dominated by the KBM with a transport ratio $\chi_e / D_e = 3/2$. \Cref{eq:deltaped_KBMtransport,eq:deltaped_KBMtransport_two} have noteworthy features: (a) there is a minimum level of transport $\langle q_e / \Gamma_e \rangle_{\mathrm{ped,3/2}} \simeq 3(1 + d)/2$ required for $\Delta_{\mathrm{ped}} > 0$, (b) if $\langle \eta_e \rangle_{\mathrm{ped} } $ is independent of $\beta_{\theta, \mathrm{ped} }$, then a fixed $\langle q_e / \Gamma_e \rangle_{\mathrm{ped}}$ value can support any pedestal width, and (c) $\Delta_{\mathrm{ped} } '$ is an increasing function of $\langle q_e / \Gamma_e \rangle_{\mathrm{ped,3/2}}$ at fixed $n_{e,\mathrm{ped}}$ but decreasing for fixed $T_{e,\mathrm{ped}}$.

In the next section, we show that in marginally stable regions close to -- but not exactly along -- the GCP, transport ratios can differ substantially.

\subsection{Transport Around GCP Marginality} \label{sec:transp_marginal}

 \begin{figure}[tb]
    \centering
    \includegraphics[width=0.9\textwidth]{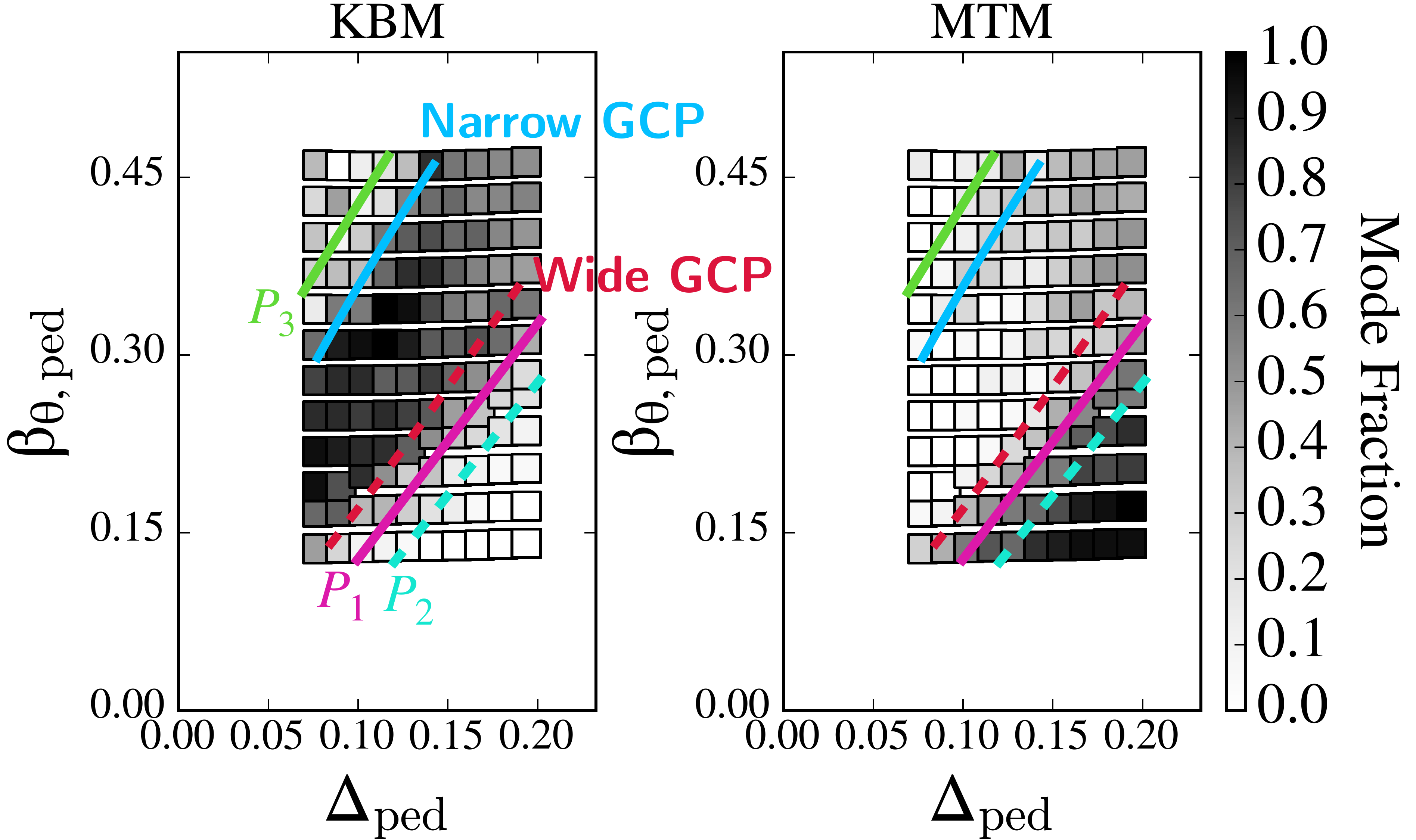}
        \caption{Paths $P_1$, $P_2$, and $P_3$ in the vicinity of the wide and narrow GCPs, plotted over KBM and MTM mode dominance fraction versus pedestal width and height for NSTX 139047 $A_2$. The mode fraction is the fraction of all linear gyrokinetic simulations across the pedestal half-width and binormal wavenumbers (here $k_y \rho_i \in [0.06, 0.12, 0.18]$) where the fastest growing mode is classified as KBM, MTM, etc.}
    \label{fig:marginality}
\end{figure}

In this section, we describe turbulent transport ratios in the vicinity of the wide and narrow GCP branches for NSTX 139047 $A_2$. The wide and narrow GCP branches can be seen as experimental bounds for $\beta_{\theta, \mathrm{ped} }$: KBM-limited pedestals may not fall exactly on the GCP branches in the inter-ELM cycle. Therefore, it is instructive to study transport not just along the GCP, but also in its vicinity.

Transport ratios change at pressures below the wide GCP scaling. Consider a path where rather than all radial locations in the pedestal half-width being unstable to KBM, only half of all radial locations are unstable -- this is shown by path $P_1$ in \Cref{fig:marginality}. Along $P_1$ there is still a substantial KBM mode fraction (roughly 1/3 of modes), so KBM transport may still be expected. Of the remaining KBMs, $\chi_e/D_e$ and $\chi_e / D_C$ both increase, meaning that electron and impurity particle transport are increasing relative to heat transport. In \Cref{tab:tab_transp_coeffs}, we summarize the KBM transport properties of $P_1$ in the `Stable, Near Wide Branch' row. Along a trajectory above the narrow BCP branch, path $P_3$ in \Cref{fig:marginality}, the ratio $\chi_e/D_e$ also increases while $\chi_e/D_C$ decreases. This indicates that KBM impurity particle transport might be more significant in the narrow than the wide branch.
\begin{table}[tb]
\centering
\caption{\label{tab:example} KBM transport coefficients in NSTX 139047 $A_2$ in different GCP regions.} %
  \begin{tabular}{| c || cc |}
    \hline
    GCP & $\chi_e / D_e$ & $\chi_e / D_C$ \\ \hline
    Strongly Unstable & $\simeq 3/2$ & $\simeq 15-25$ \\ \hline
    Stable, Near Wide Branch & $\simeq 3/2 - 5/2$ & $\simeq 25-40$ \\ \hline
    Stable, Near Narrow Branch & $\simeq 3/2 - 5$ & $\simeq 5-15$ \\ \hline
  \end{tabular}
\label{tab:tab_transp_coeffs}
\end{table}

Non-KBM transport is likely increasingly important further away from the GCP. Consider the path $P_2$ in \Cref{fig:marginality} where the pedestal is at a pressure far below the GCP. Here, the KBM is close to marginally stable at all pedestal flux surfaces. Shown in \Cref{fig:marginality}(b), the MTM has a significant mode fraction along $P_2$ for the $k_y \rho_i$ values we simulate, indicating significant electron heat transport. Given that KBM is subdominant along $P_2$, another mechanism such as TEM or neoclassical physics is required for particle transport.

\subsection{Aspect-Ratio Dependence}

In this section, we show how transport near the GCP varies across aspect-ratio.

\begin{figure*}[tb]
    \centering
    \begin{subfigure}[t]{0.47\textwidth}
        \centering
        \includegraphics[width=\textwidth]{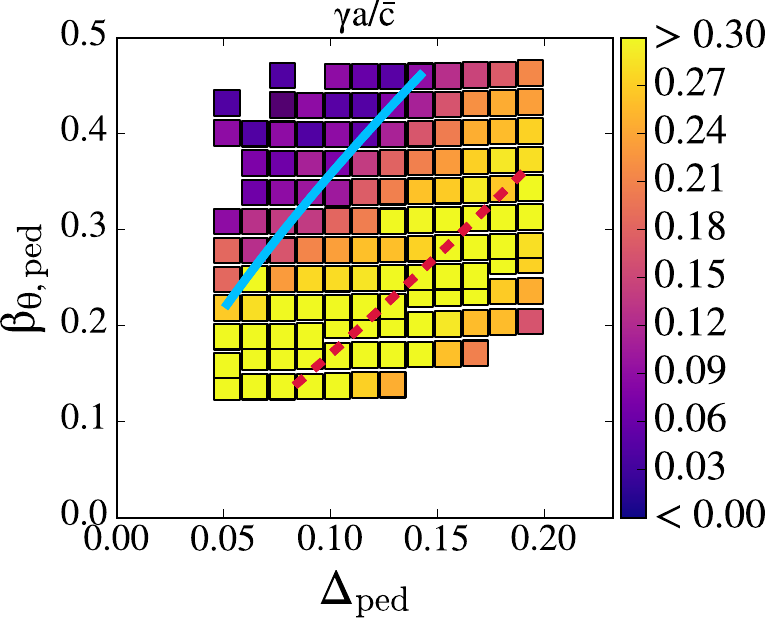}
        \caption{Aspect-ratio = 2.0}
    \end{subfigure}
    ~
    \begin{subfigure}[t]{0.47\textwidth}
        \centering
        \includegraphics[width=\textwidth]{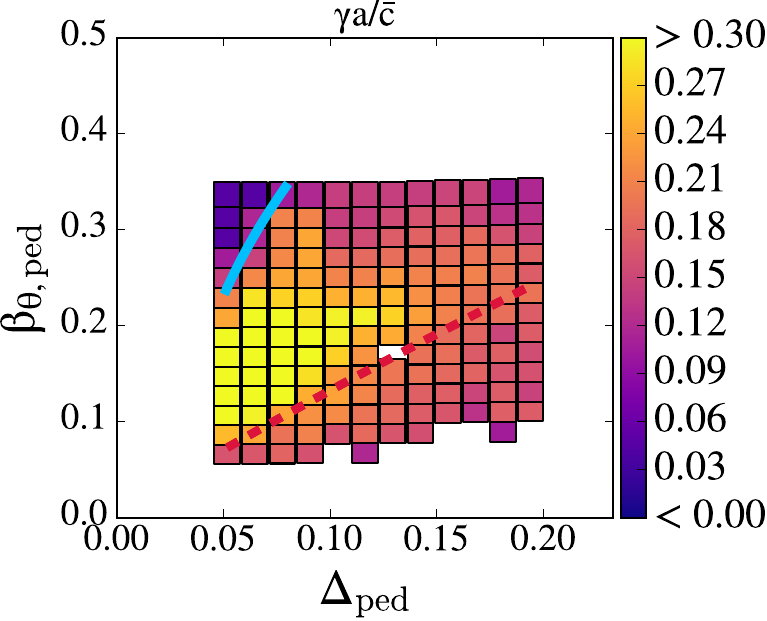}
        \caption{Aspect-ratio = 2.9}
    \end{subfigure}
    \caption{KBM linear growth rates averaged over the pedestal half-width and all simulated wavenumbers versus $\Delta_{\mathrm{ped} }$ and $\beta_{\theta,\mathrm{ped} }$ for two aspect-ratios: NSTX 139047 $A_2$ (a) and $A_4$ (b).}
    \label{fig:marginality_growth_rates_AR}
\end{figure*}

Along the wide GCP, $\chi_e / D_e$ along the wide GCP branch is larger at is higher-A ($\chi_e / D_e \simeq 2-2.5$) than at lower-A ($\chi_e / D_e \simeq 1.5$). At lower-A, any remaining KBM instabilities also have a higher growth rate than at higher-A along the wide GCP branch, shown in \Cref{fig:marginality_growth_rates_AR} where we plot the average growth rate $\gamma$ of KBM modes for NSTX 139047 $A_2$ (A = 2.0) and $A_4$ (A = 2.9). Here, $\gamma$ is normalized by the minor radius $a$ and sound speed $\bar{c}$. This indicates that the more unstable KBMs along the lower-A wide GCP branch have transport properties expected for an MHD mode $\chi_e / D_e = 3/2$. At lower growth rates close to KBM marginality, the higher-A KBM has modified transport properties where kinetic effects appear to be particularly important. However, along the narrow GCP branch for both lower and higher aspect-ratio in \Cref{fig:marginality_growth_rates_AR}, the KBM growth rate decreases sharply around the narrow GCP boundary, indicating that KBM transport changes quickly across the boundary.

\begin{figure*}[tb]
    \centering
    \begin{subfigure}[t]{0.44\textwidth}
        \centering
        \includegraphics[width=\textwidth]{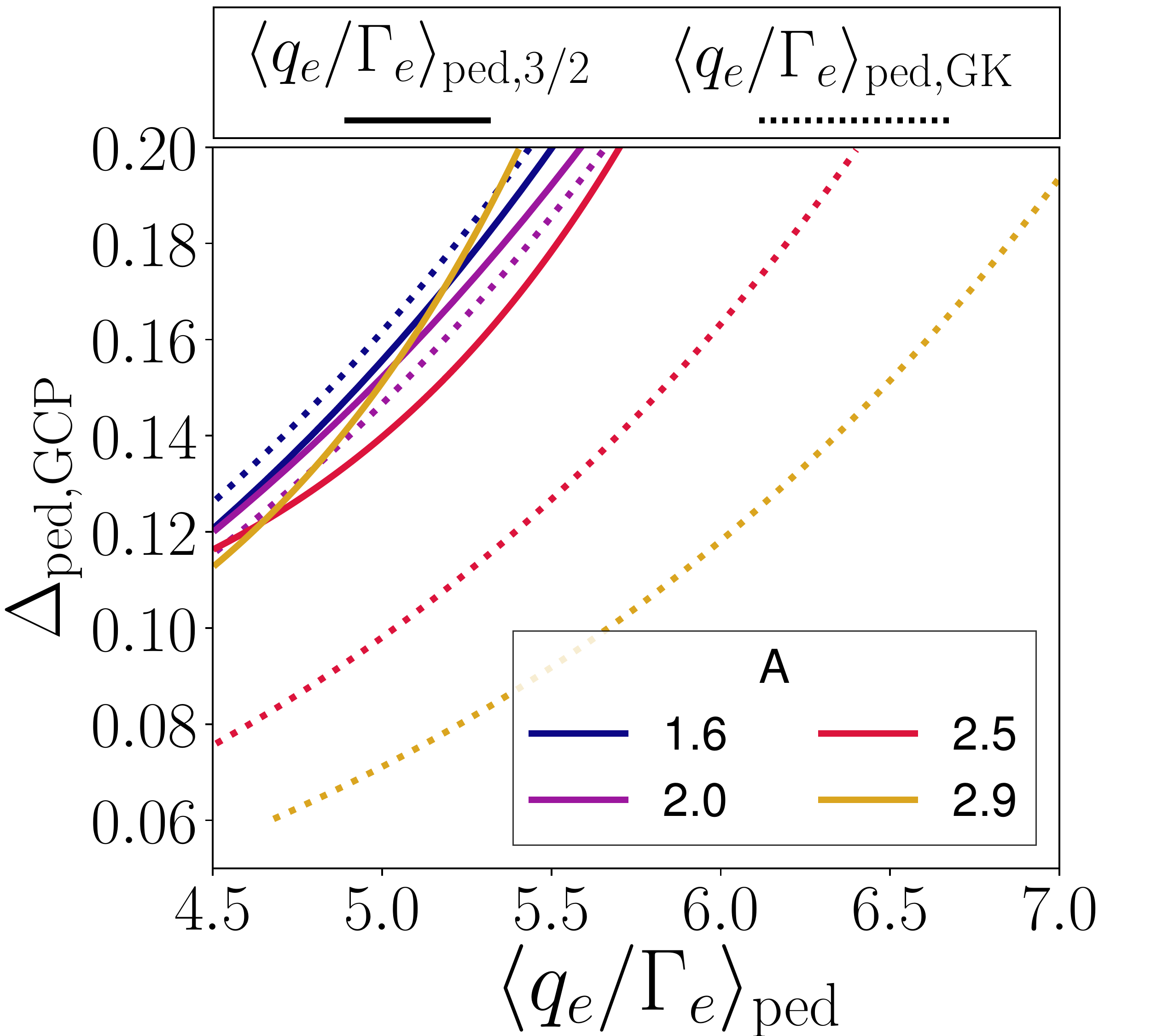}
        \caption{Width versus $\langle q_e/\Gamma_e \rangle_{\mathrm{ped} }$ for model with $\chi_e/D_e = 3/2$ and KBM from gyrokinetic simulations.}
    \end{subfigure}
    ~
    \begin{subfigure}[t]{0.52\textwidth}
        \centering
        \includegraphics[width=\textwidth]{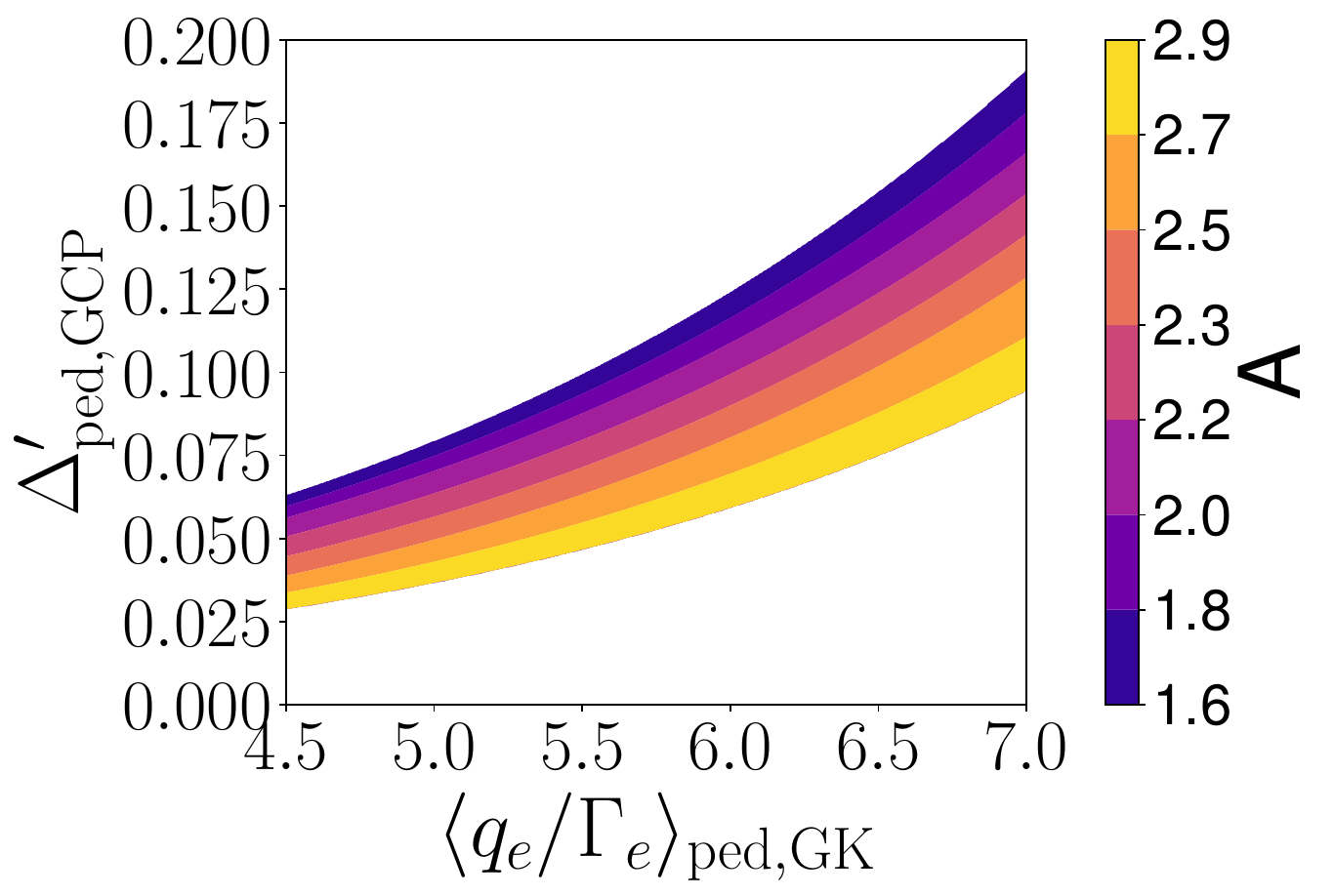}
        \caption{Width derivative versus $\langle q_e/\Gamma_e \rangle_{\mathrm{ped, GK} }$ for KBM from gyrokinetic simulations.}
    \end{subfigure}
    \caption{Pedestal width (a) and its derivative (b) versus $\langle q_e/\Gamma_e \rangle_{\mathrm{ped} }$ for NSTX 139047 $A_2$ at fixed $n_{e,\mathrm{ped}}$.}
    \label{fig:etaescaling_heat_aspectratio}
\end{figure*}

The differing KBM transport coefficients across aspect-ratio along the GCP affects the width-transport scaling. In \Cref{fig:etaescaling_heat_aspectratio}(a), we show the difference between a prediction for $\Delta_{\mathrm{ped,GCP} }$ using the $\chi_e / D = 3/2$ model and the data from gyrokinetic simulations. At low $A$ values, $\Delta_{\mathrm{ped,GCP} }$ is the same for both $\langle q_e/\Gamma_e \rangle_{\mathrm{ped,3/2} }$ and $\langle q_e/\Gamma_e \rangle_{\mathrm{ped,GK} }$, whereas at higher values there is a very large discrepancy. This difference comes from the increased $\chi_e/D_e$ value of the KBM for larger $A$ values. We also plot the derivative $\Delta_{\mathrm{ped,GCP} } '$ in \Cref{fig:etaescaling_heat_aspectratio}(b). The quantity $\Delta_{\mathrm{ped,GCP} } '$ has an aspect-ratio dependence --- the smaller $A$, the more sensitive $\Delta_{\mathrm{ped,GCP} }$ is to $\langle q_e/\Gamma_e \rangle_{\mathrm{ped,GK} }$. Writing the derivative of $\Delta_{\mathrm{ped,GCP} }$ in the form of \Cref{eq:width_scaling_general} with the diffusivity $\chi_e / D_e$ included explicitly,
\begin{equation}
\begin{aligned}
& \Delta_{\mathrm{ped,GCP} } ' = \frac{d \Delta_{\mathrm{ped,GCP} }}{d \langle q_e/\Gamma_e \rangle_{\mathrm{GCP} } }  \\
& = \left( \frac{D_e}{\chi_e} \frac{1}{a_1 D^{\alpha_1}}  \right)^{1/(\alpha_1 \alpha_2)}  \left( \frac{ \frac{D_e}{\chi_e} \left( \langle \frac{q_e}{\Gamma_e} \rangle_{\mathrm{ped} } - \frac{3}{2} \right) - d }{a_1 \left( D \right)^{\alpha_1}}  \right)^{-1+1/(\alpha_1 \alpha_2) },
\end{aligned}
\end{equation}
one sees that increased $\chi_e/D_e$ values generally decrease $\Delta_{\mathrm{ped,GCP} }$ and $\Delta_{\mathrm{ped,GCP} } '$.

\subsection{Radial Dependence}

In this section, we describe briefly the radial dependence of gyrokinetic instabilities in pedestal width and height space for NSTX 139047 $A_2$. We study the `average' radial location of an instability between $\psi_{1/4} = \psi_{\mathrm{mid} } - \Delta_{\mathrm{ped} }/4$ and $\psi_{3/4} = \psi_{\mathrm{mid} } + \Delta_{\mathrm{ped} }/4$ using the quantity
\begin{equation}
Y = 4 \frac{\psi - \psi_{\mathrm{mid} }}{\Delta_{\mathrm{ped} }},
\end{equation}
such that $Y_{1/4} = Y(\psi_{1/4}) = -1$ and $Y_{3/4} = Y(\psi_{3/4}) = 1$. Averaging  $Y$ for a given mode,
\begin{equation}
\langle Y \rangle_{\mathrm{mode} } \equiv \frac{\sum_{k_y} \int_{-1}^{1} Y \; \mathrm{mode}(Y, k_y) \; dY}{\sum_{k_y,} \int_{-1}^{1} dY},
\label{eq:Y_radav}
\end{equation}
gives the average $Y$ radial value across the pedestal half-width. Here, mode=1 if the mode is present and mode=0 if not. For example, if KBM were unstable at all radial and $k_y \rho_i$ values then $\langle Y \rangle_{\mathrm{KBM} } = 0$, indicating a uniform radial distribution across the pedestal half-width relative to the pedestal mid-radius.

 \begin{figure}[tb]
    \centering
    \includegraphics[width=\textwidth]{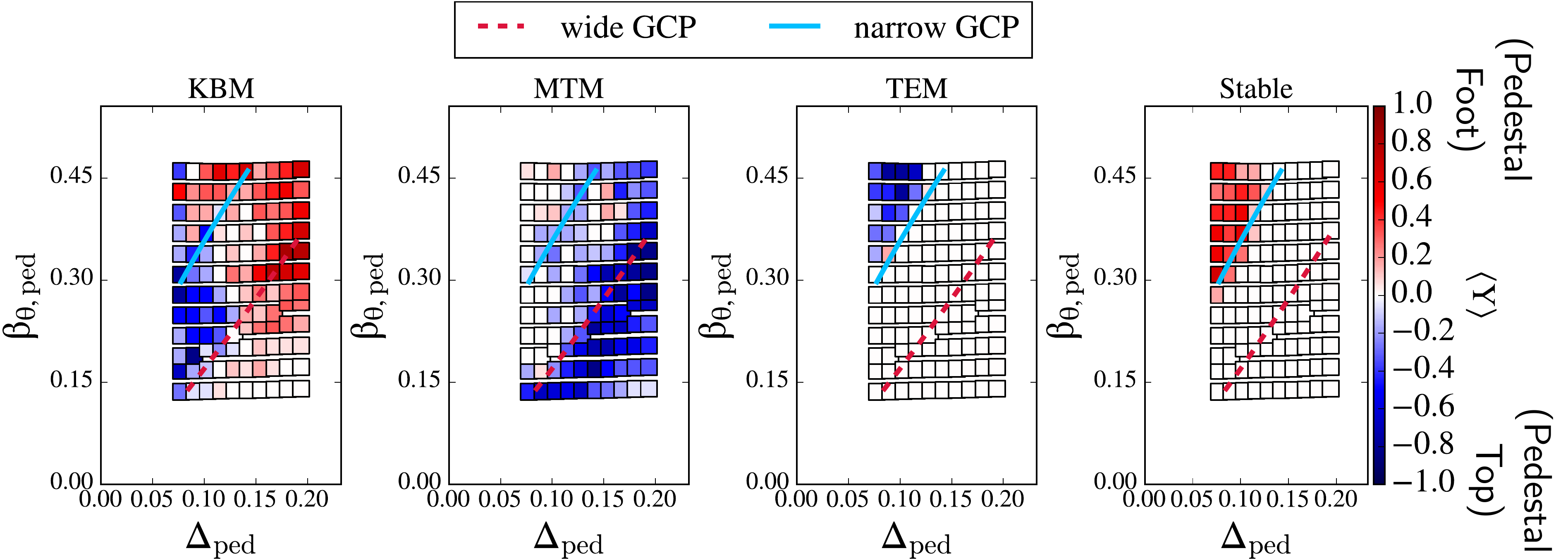}
        \caption{Average radial location $\langle Y \rangle $ defined in \Cref{eq:Y_radav} of mode type across the pedestal for NSTX 139047 $A_2$ across wavenumbers $k_y \rho_i \in [0.06, 0.12, 0.18]$.}
    \label{fig:modetype_radialfraction}
\end{figure}

\Cref{fig:modetype_radialfraction}(a) shows that moving along the wide and narrow GCP, initially at small widths the KBM is more common at the pedestal top, but as the pedestal width increases, the KBM is more common around the pedestal foot. This is likely due to the magnetic shear profiles changing as the bootstrap current density increases, shifting the relatively higher magnetic shear (and therefore destabilizing) regions from the pedestal top to the pedestal foot at higher $\Delta_{\mathrm{ped} }$, $\beta_{\theta, \mathrm{ped}}$ values. For MTMs in \Cref{fig:modetype_radialfraction}(b), the instability is most common near the pedestal top, consistent with other works \cite{Dickinson2012, Dickinson2013, Hatch2016}. \Cref{fig:modetype_radialfraction}(c) shows that the TEM only appears in KBM second stability and is also more prevalent at the pedestal top, also seen in other works \cite{Fulton2014,Guttenfelder2021}. The lack of TEM instability below the wide GCP branch at lower gradients might result from the relatively narrow $k_y \rho_i \in [0.06, 0.12, 0.18]$ range we include in simulations for this work -- because steeper gradients tend to destabilize modes at lower $k_y \rho_i$ values \cite{Barnes2011,Parisi2022}, we likely require higher $k_y \rho_i$ values for the pedestals below the wide GCP branch to find TEM instability. Despite the ubiquity of ETG instability in the pedestal \cite{Hatch2016, Kotschenreuther2019, Chapman2022, Parisi2020}, we find relatively little dominant ETG modes for $k_y \rho_i \in [0.06, 0.12, 0.18]$, likely because we are again only simulating binormal wavenumbers that are too low. Finally, \Cref{fig:modetype_radialfraction}(d) shows that in KBM second stability, much of the pedestal foot is stabilized, which is the only location in $\beta_{\theta, \mathrm{ped} }$, $\Delta_{\mathrm{ped} }$ space where stabilization occurs for $k_y \rho_i \in [0.06, 0.12, 0.18]$.

\section{Flow Shear Along the Gyrokinetic Critical Pedestal} \label{sec:flowshear}

In this section, we study the efficacy of pedestal flow shear for two NSTX discharges. The radial shear in plasma rotation $\Omega_{\zeta}$ is important in the L-H transition \cite{Groebner1990,Rogers1998,Cavedon2020} and inter-ELM pedestal dynamics \cite{Schirmer2006,Snyder2007, Hatch2017, Yan2011,Chen2017b,Barada2018}, where $\Omega_{\zeta}$ is the toroidal angular rotation frequency. One key assumption in the EPED model is that flow shear at the pedestal top suppresses turbulence \cite{Snyder2009} allowing the pedestal to broaden. We examine this assumption by comparing flow shear in a standard (139047) and wide (132588) NSTX pedestal along the GCP. %

\begin{figure*}[tb]
    \centering
    \begin{subfigure}[t]{0.48\textwidth}
        \centering
        \includegraphics[width=\textwidth]{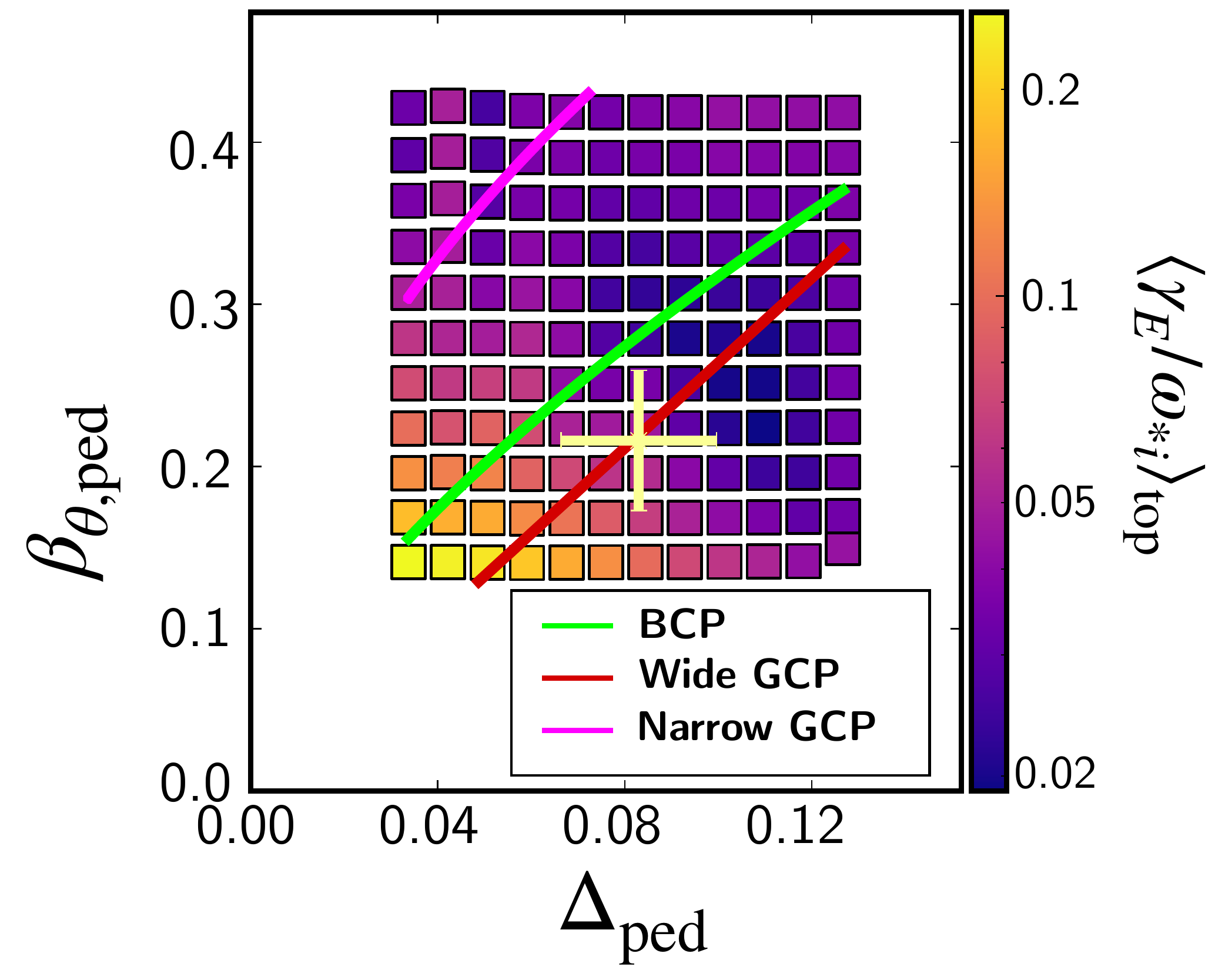}
        \caption{NSTX 139047.}
    \end{subfigure}
    \begin{subfigure}[t]{0.48\textwidth}
        \centering
        \includegraphics[width=\textwidth]{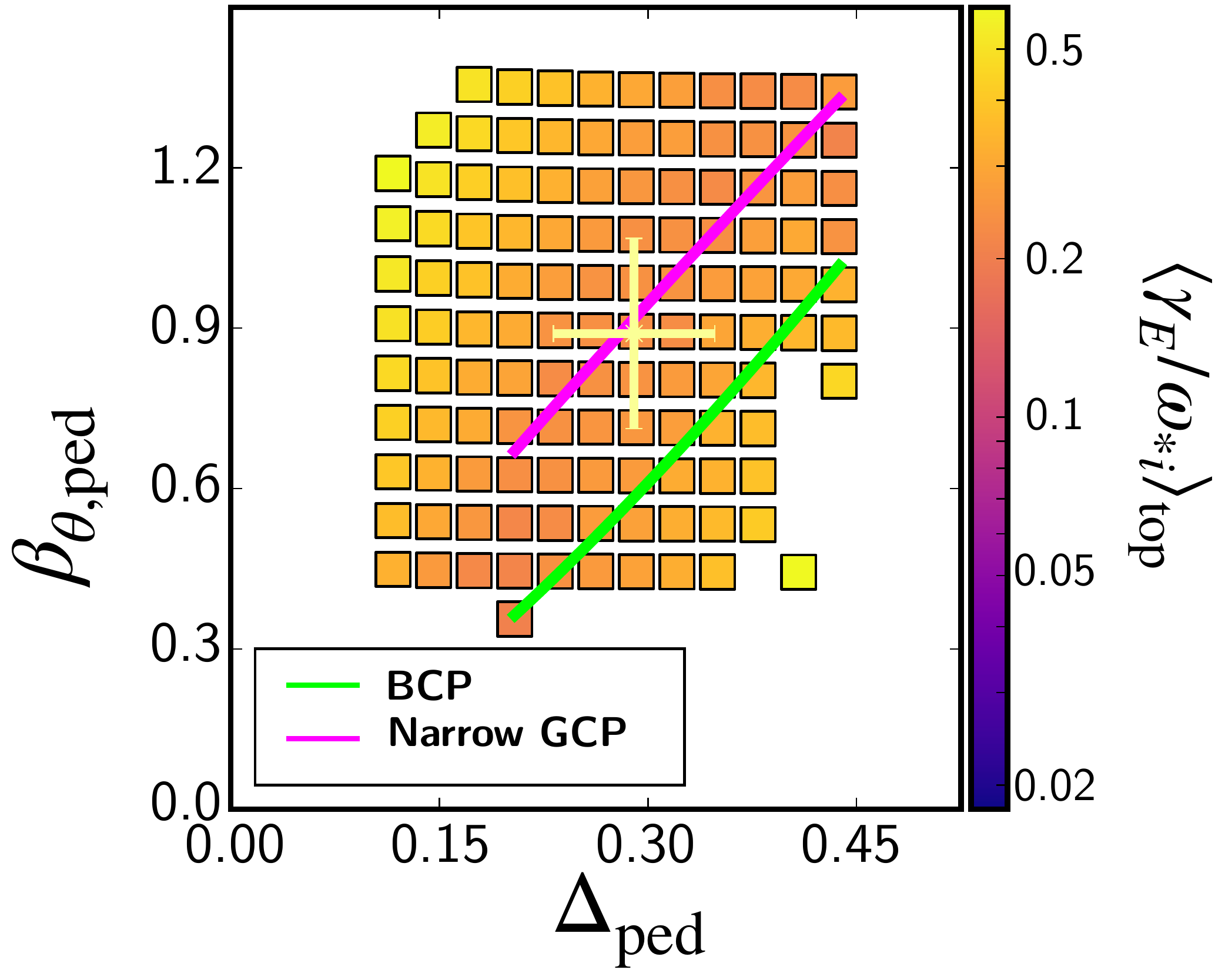}
        \caption{NSTX 132588.}
    \end{subfigure}
    \caption{Pedestal-top-averaged flow shear $\gamma_E/\omega_{*i}$ (see \Cref{eq:gammaE,eq:omegastar,eq:flow_shear_ped_top}) for NSTX 139047 and 132588, with the BCP, GCP, and equilibrium point (yellow marker).}
    \label{fig:flowshearfreq}
\end{figure*}

Balancing the radial electric field with the diamagnetic flow \cite{Hatch2016, Parisi2020}, we approximate the flow shear rate $\gamma_E = (r/q) d \Omega_{\zeta}/ dr$ as
\begin{equation}
    \gamma_E \approx \frac{r}{q} \frac{\partial }{\partial r} \left( \frac{dp_i}{d\psi} \frac{c}{Z_i e n_i} \right).
    \label{eq:gammaE}
\end{equation}
As a rough estimate of flow shear efficacy, we compare the flow shear frequency with the linear drive frequency
\begin{equation}
    \omega_{*i} = v_{ti} \sum_s \left( \frac{1}{L_{Ts}} + \frac{1}{L_{ns}} \right),
    \label{eq:omegastar}
\end{equation}
which is typically comparable to the linear instability growth rate. In \Cref{fig:flowshearfreq}(a), we plot $\gamma_E / \omega_{*i}$ averaged over the pedestal top,
\begin{equation}
\left\langle \frac{\gamma_E}{\omega_{*i} }  \right\rangle_{\mathrm{top} } = (4/\Delta_{\mathrm{ped}}) \int_{\psi_{\mathrm{ped}}}^{\psi_{1/4 } } \frac{q_e}{\Gamma_e} \; d \psi,
\label{eq:flow_shear_ped_top}
\end{equation}
versus pedestal width and height for NSTX 139047. Here, $\psi_{1/4} = \psi_{\mathrm{mid} } - \Delta/4$ and $\psi_{\mathrm{ped} } = \psi_{\mathrm{mid} } - \Delta/2$. At the pedestal top, larger values of $\left\langle \gamma_E/\omega_{*i} \right\rangle_{\mathrm{top} }$ suppress turbulence and may facilitate the pedestal's radially inward propagation \cite{Snyder2009}, although too large values may cause other instabilities \cite{Catto1973,Newton2010,Schekochihin2012}. Moving along the wide GCP and BCP starting from small widths and heights, $\left\langle \gamma_E/\omega_{*i} \right\rangle_{\mathrm{top} }$ is a rapidly decreasing function. The marked equilibrium point for NSTX 139047 in \Cref{fig:flowshearfreq}(a) is at a location where $\left\langle \gamma_E/\omega_{*i} \right\rangle_{\mathrm{top} }$ approaches a minimum -- moving further along the wide GCP by increasing the width-height may give pedestals with insufficient pedestal top flow shear to permit an increase in pedestal width.

For comparison, we consider pedestal flow shear for ELM-free wide pedestal NSTX discharge 132588. In \Cref{fig:flowshearfreq}(b), we plot $\left\langle \gamma_E/\omega_{*i} \right\rangle_{\mathrm{top} }$. Compared with NSTX 139047 in \Cref{fig:flowshearfreq}(a), the wide pedestal has very strong pedestal top flow shear. This strong flow shear indicates that this discharge may have sufficient flow shear to allow the pedestal width to grow to very large values.

We emphasize that using the ratio $\gamma_E/\omega_{*i}$ is only heuristic, and actual gyrokinetic simulations are required to definitively determine flow shear efficacy across the $\beta_{\theta, \mathrm{ped} }$, $\Delta_{\mathrm{ped} }$ space. Additionally, given that flow shear can also affect peeling-ballooning stability \cite{Xi2012}, determining the effect of pedestal flow shear on both microstability and macrostability is necessary to understand its role in pedestal evolution.

\section{Combined Stability, Flow Shear, and Transport Constraints} \label{sec:combined_constraints}

\begin{figure}[tb]
    \centering
    \includegraphics[width=0.8\textwidth]{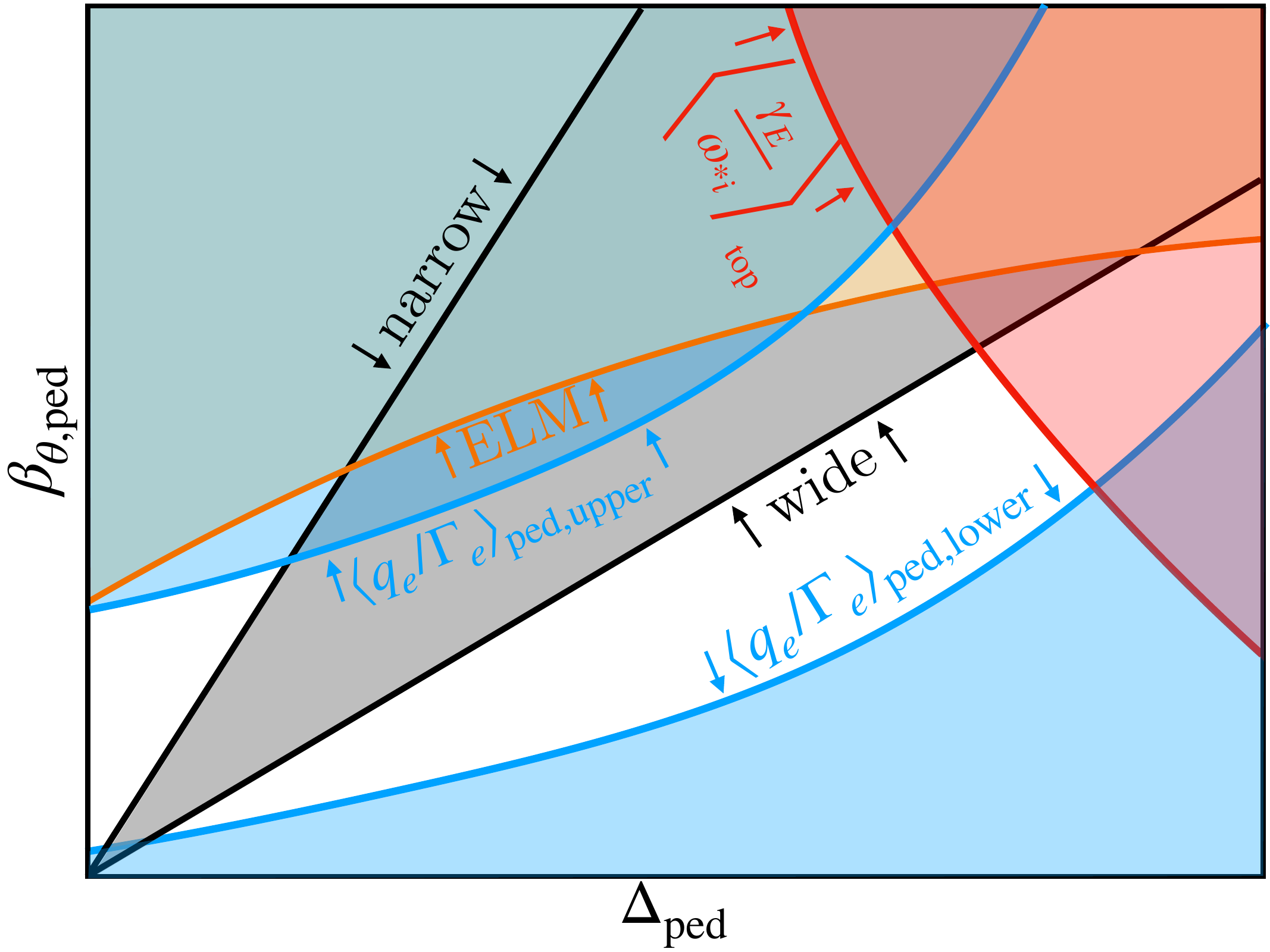}
        \caption{Combination of stability, flow shear, transport, and ELM constraints. Arrows indicate the direction that the constraint acts on in $\beta_{\theta, \mathrm{ped} }$, $\Delta_{\mathrm{ped} }$ space. The translucent color blocks indicate the excluded regions for a constraint curve of the same color. The only accessible pedestal regions are those without any constraints. In this example, for the transport constraints $\langle q_e / \Gamma_e \rangle_{\mathrm{ped,lower} } $ and $\langle q_e / \Gamma_e \rangle_{\mathrm{ped,upper} }$, we have assumed that $\eta_e$ increases with $\beta_{\theta, \mathrm{ped} }$.}
    \label{fig:combined_constraints}
\end{figure}

In this section, we discuss the combination of three pedestal constraints discussed in this paper: KBM stability, flow shear, and transport. Due to its importance as a saturation mechanism, we also include an ELM constraint \cite{Snyder2009}.

For KBM stability, the Gyrokinetic Critical Pedestal gives the wide and narrow branches, indicated by lines labeled wide and narrow in \Cref{fig:combined_constraints}. According to kinetic-ballooning-mode stability, the pedestal can only manifest at pressures above the narrow branch and pressures below the wide branch. Arrows in \Cref{fig:combined_constraints} along the wide and narrow branches indicate the direction in which the constraint applies, ruling out pedestal access to these excluded regions of $\beta_{\theta, \mathrm{ped} }$ and $\Delta_{\mathrm{ped} }$.

For flow shear, we show a single constraint given by $\left\langle \gamma_E/\omega_{*i} \right\rangle_{\mathrm{top} }$ in \Cref{fig:combined_constraints}: values below a critical value indicate that the pedestal top flow shear is no longer sufficiently strong to allow the pedestal to widen. From the results in \Cref{sec:flowshear} we graph the flow shear constraint as $\beta_{\theta, \mathrm{ped} } \sim 1/\Delta_{\mathrm{ped} }$.

For transport, we plot two constraints in \Cref{fig:combined_constraints}: $\left\langle q_e / \Gamma_e \right\rangle_{\mathrm{ped,lower}}$ and $\left\langle q_e / \Gamma_e \right\rangle_{\mathrm{ped,upper}}$, which are the minimum and maximum values of $\left\langle q_e / \Gamma_e \right\rangle_{\mathrm{ped}}$ that the particle and heat sources might permit to support a given pedestal profile. In steady state for heat and particle sources $P_e$ and $S_e$
\begin{equation}
\left\langle \frac{q_e}{\Gamma_e} \right\rangle_{\mathrm{ped} }= \frac{\int P_e dV_{\mathrm{half-width}} }{\int S_e dV_{\mathrm{half-width}}},
\label{eq:eta_flux_ratio_av_sources}
\end{equation}
where the volume integral is taken over the pedestal half-width. For the example in \Cref{fig:combined_constraints}, we have assumed that $\eta_e$ increases with $\beta_{\theta, \mathrm{ped} }$ and hence $\left\langle q_e / \Gamma_e \right\rangle_{\mathrm{ped}}$ also increases with $\beta_{\theta, \mathrm{ped} }$ according to \Cref{eq:eta_flux_ratio_av}. If however, $\eta_e$ were to decrease with $\beta_{\theta, \mathrm{ped} }$, the transport constraints in \Cref{fig:combined_constraints} would change significantly.

Finally, we plot a peeling-ballooning ELM constraint in \Cref{fig:combined_constraints} assuming $\Delta_{\mathrm{ped} } \sim \left( \beta_{\theta, \mathrm{ped} } \right)^{4/3}$ \cite{Snyder2009}. Pedestal pressures above the ELM constraint are inaccessible.

\Cref{fig:combined_constraints} shows that after considering stability, pedestal top flow shear, and transport, only a relatively narrow window of $\beta_{\theta, \mathrm{ped} }$, $\Delta_{\mathrm{ped} }$ space remains for a viable pedestal. In the heuristic example in \Cref{fig:combined_constraints}, moving along the wide GCP branch, the pedestal growth is eventually stopped by the flow shear constraint. Moving along the narrow GCP branch, the pedestal growth is stopped by the upper transport constraint.

\section{Summary}

In this work, we described how a linear gyrokinetic threshold model combined with self-consistent equilibrium variation gives a pedestal width-height scaling expression that depends strongly on aspect-ratio and plasma shaping. These shaping and aspect-ratio scans were performed on an NSTX equilibrium with self-consistent equilibrium reconstruction. The Gyrokinetic Critical Pedestal scaling for the wide branch kinetic-ballooning-mode pedestals with shaping and aspect-ratio dependence in this NSTX equilibrium is $\Delta_{\mathrm{ped}} = 0.92 A^{1.04} \kappa^{-1.24} 0.38^{\delta} \left( \beta_{\theta,\mathrm{ped}}\right)^{1.05}$. It is noteworthy that the width-height scaling has a strong dependence on shaping and aspect-ratio for both the wide and barrow GCP branches \cite{Parisi_arxiv2023_2} --- this dependence might not have been reported experimentally due to the relatively narrow range of aspect-ratio, triangularity, and elongation values routinely examined compared with those in the parameter scan in this work. In future work, it is important to study the effect of shaping on both kinetic-ballooning \cite{Parisi_arxiv2023_2} and peeling-ballooning \cite{Snyder2015,Merle2017,Snyder2019} stability to definitively find attractive pedestal scenarios. %

We demonstrated that whether pedestal height $\beta_{\theta, \mathrm{ped}}$ is varied with fixed $n_{e,\mathrm{ped} }$ or $T_{e,\mathrm{ped} }$ changes heat and particle transport along the Gyrokinetic Critical Pedestal significantly. In the wide Gyrokinetic Critical Pedestal branch for fixed $n_{e,\mathrm{ped} }$ we find that $\Delta_{\mathrm{ped} } = 0.028 \left(q_e/\Gamma_e - 1.7 \right)^{1.5} \sim \left( \eta_e \right)^{1.5}$ and for fixed $T_{e,\mathrm{ped} }$ we find $\Delta_{\mathrm{ped} } = 0.31 \left(q_e/\Gamma_e - 4.7 \right)^{0.85} \sim \left( \eta_e \right)^{0.85}$. Thus, during the inter-ELM buildup, not only do relative contributions to $\beta_{\theta, \mathrm{ped}}$ from temperature and density have a big effect on the width-height scaling, but also the relative transport through particle and heat channels. While our linear gyrokinetic model has no information about plasma sources, more sophisticated approaches such as transport solvers \cite{Candy2009,Barnes2010trinity,Siena2022} will be sensitive to KBM transport around the Gyrokinetic Critical Pedestal.

In the vicinity of the narrow and wide Gyrokinetic Critical Pedestal branches where KBM becomes marginally stable, the KBM turbulent transport ratios such as $\chi_e/D_e$ and $\chi_e/D_Z$ vary significantly. This also has implications for reduced transport models in the pedestal, given that KBM can produce significant particle and heat flux.

We analyzed the role of flow shear for two NSTX pedestal discharges, finding that the width and height of an ultra-wide-pedestal ELM-free NSTX discharge had much stronger flow shear at the pedestal top, possibly permitting its pedestal to grow to a very large width. This indicates that the flow shear can introduce an additional constraint on the pedestal width and height evolution, and might be an important pedestal growth saturation mechanism.

\section{Acknowledgements}

We are grateful for conversations with E. A. Belli, J. Candy, S. C. Cowley, D. Dickinson, W. Dorland, R. Maingi, F. I. Parra, M. J. Pueschel, O. Sauter, P. B. Snyder, G. M. Staebler, and H. R. Wilson. This work was supported by the U.S. DoE under contract numbers DE-AC02-09CH11466, DE-SC0022270, DE-SC0022272, and the DoE Early Career Research Program. The US Government retains a non-exclusive, paid-up, irrevocable, world-wide license to publish or reproduce the published form of this manuscript, or allow others to do so, for US Government purposes.

\appendix

\section{$\langle \eta_e \rangle_{\mathrm{ped} }$} \label{app:eta_rescaling}

Here we calculate useful expressions for $\eta_e$ and $\langle \eta_e \rangle_{\mathrm{ped} }$. Using new radial coordinates $F_{T_e}$ and $F_{n_e}$, 
\begin{equation}
F_{T_e} = 2 \frac{ \psi - \psi_{\mathrm{ mid, T_e}} } {\Delta_{\mathrm{ped},T_e}}, \;\;\; F_{n_e} = 2 \frac{ \psi - \psi_{\mathrm{ mid, n_e}} } {\Delta_{\mathrm{ped},n_e}},
\end{equation}
the electron temperature profile is
\begin{equation}
T_e (F_{T_e}) = T_{e0}\left(\tanh(2) - \tanh(F_{T_e})\right) + T_{e,\mathrm{sep}},
\end{equation}
and the density profile is
\begin{equation}
n_e (F_{n_e}) = 	n_{e0}\left(\tanh(2) - \tanh(F_{n_e})\right) + n_{e,\mathrm{sep}},
\end{equation}
which gives an expression for $\eta_e$,
\begin{equation}
\eta_e = \frac{L_{n_e}}{L_{T_e}}  = \frac{\Delta_{\mathrm{ped},n_e }}{\Delta_{\mathrm{ped},T_e }} \frac{\mathrm{sech}(F_{T_e})^2 [(n_{e,\mathrm{sep}}/n_{e0}) + \tanh(2) - \tanh(F_{n_e})]}{\mathrm{sech}(F_{n_e})^2[(T_{e,\mathrm{sep}}/T_{e0}) + \tanh(2) - \tanh(F_{T_e})]}.
\label{eq:general_etae}
\end{equation}

\subsection{Width-Independent Limit}

In the limit where $F_{T_e} = F_{n_e} = F$ and $\Delta_{\mathrm{ped},T_e } = \Delta_{\mathrm{ped},n_e }$,
\begin{equation}
\eta_e = \frac{ [\overline{n} + \tanh(2) - \tanh(F)]}{ [\overline{T} + \tanh(2) - \tanh(F)]},
\label{eq:etae_delta_ind}
\end{equation}
where $\overline{n} = n_{e,\mathrm{sep}}/n_{e0}$ and $\overline{T} = T_{e,\mathrm{sep}}/T_{e0}$.  Averaging $\eta_e$ over the past half-width gives
\begin{equation}
\begin{aligned}
& \langle \eta_e \rangle_{\mathrm{ped} } = 2 \int_{\psi_{\mathrm{mid} } - \Delta_{\mathrm{ped} }/4}^{\psi_{\mathrm{mid} } + \Delta_{\mathrm{ped} }/4} \eta_e d\psi / \Delta_{\mathrm{ped} } = \int_{F = -1/2}^{F=1/2} \eta_e d F = \\
& \frac{\overline{n}(\overline{T} -1) - \overline{ T} + 2e^4 (\overline{n} \overline{ T} - 2 )  + e^8 (\overline{n} + \overline{T}  + \overline{ n} \overline{ T} ) + (1+e^4)^2 (\overline{n} - \overline{ T} ) X }{(\overline{T}(1+e^4) -2)(\overline{T} + e^4 (\overline{ T} + 2) ) },
\label{eq:etae_average_delta_ind}
\end{aligned}
\end{equation}
where
\begin{equation}
X = \ln \frac{\overline{T} + \tanh(2) - \tanh(1/2)}{\overline{T} + \tanh(2) + \tanh(1/2)}.
\end{equation}
The quantities $\eta_e$ and $\langle \eta_e \rangle_{\mathrm{ped} }$ in \Cref{eq:etae_delta_ind,eq:etae_average_delta_ind} are independent of the pedestal width.
\printbibliography

\end{document}